\documentclass{aa}  

\usepackage{graphicx}
\usepackage{txfonts}
\usepackage{hyperref}
\usepackage{booktabs}
\usepackage{bm}
\usepackage[normalem]{ulem}
\usepackage[dvipsnames]{xcolor}
\usepackage{nicefrac}

\newcommand{\tauav}{\tau_\mathrm{av}}
\newcommand{\tauconv}{\tau_\mathrm{conv}}
\newcommand{\vw}[1]{\overline{#1}}
\newcommand{\mw}[1]{\widetilde{#1}}
\newcommand{\code}[1]{\texttt{#1}}
\newcommand{\ppmstar}{\code{PPMSTAR}}
\newcommand{\flash}{\code{FLASH}}
\newcommand{\music}{\code{MUSIC}}
\newcommand{\prompi}{\code{PROMPI}}
\newcommand{\slh}{\code{SLH}}
\newcommand{\numcodes}{five}

\setcitestyle{citesep={,}}

\begin{document} 
    \title{Dynamics in a stellar convective layer and at its boundary:
    Comparison of five 3D hydrodynamics codes}
    \titlerunning{Comparison of five 3D hydrodynamics codes}

    \author{
        R.~Andrassy\inst{1}\and
        J.~Higl\inst{1}\and
        H.~Mao\inst{2}\and
        M.~Moc\'ak\inst{3}\and
        D.~G.~Vlaykov\inst{4}\and
        % Each code's representatives above. The rest is alphabetical.
        W.~D.~Arnett\inst{5}\and
        I.~Baraffe\inst{4,6}\and 
        S.~W.~Campbell\inst{7,8}\and
        T.~Constantino\inst{4}\and 
        P.~V.~F.~Edelmann\inst{9}\and
        T.~Goffrey\inst{10}\and
        T.~Guillet\inst{4}\and 
        F.~Herwig\inst{11,12}\and
        R.~Hirschi\inst{3,13}\and
        L.~Horst\inst{1}\and
        G.~Leidi\inst{1,14}\and
        C.~Meakin\inst{5,15}\and
        J.~Pratt\inst{16}\and
        F.~Rizzuti\inst{3}\and
        F.~K.~R{\"o}pke\inst{1,17}\and
        P.~Woodward\inst{2,12}}
    \institute{
        Heidelberger Institut f{\"u}r Theoretische Studien,
        Schloss-Wolfsbrunnenweg 35, D-69118 Heidelberg, Germany\\
        \email{robert.andrassy@h-its.org}
        \and
        LCSE and Department of Astronomy, University of Minnesota, Minneapolis,
        MN 55455, USA
        \and
        Astrophysics Group, Keele University, Keele, Staffordshire ST5 5BG, UK
        \and 
        Physics and Astronomy, University of Exeter, Exeter, EX4 4QL, United
        Kingdom
        \and
        Steward Observatory, University of Arizona, 933 N.~Cherry Avenue, Tucson
        AZ 85721, USA
        \and
        \'Ecole Normale Sup\'erieure, Lyon, CRAL (UMR CNRS 5574), Universit\'e
        de Lyon, France
        \and
        School of Physics and Astronomy, Monash University, Victoria 3800,
        Australia
        \and
        ARC Centre of Excellence for All Sky Astrophysics in Three Dimensions
        (ASTRO-3D), Australia
        \and
        X Computational Physics (XCP) Division and Center for Theoretical
        Astrophysics (CTA), Los Alamos National Laboratory, Los Alamos, NM
        87545, USA
        \and
        Centre for Fusion, Space and Astrophysics,  Department of Physics,
        University of Warwick, Coventry, CV4 7AL, UK
        \and
        Department of Physics and Astronomy, University of Victoria, Victoria,
        BC, V8W 2Y2, Canada
        \and
        Joint Institute for Nuclear Astrophysics -- Center for the Evolution of
        the Elements, USA
        \and
        Kavli Institute for the Physics and Mathematics of the Universe (WPI),
        University of Tokyo, 5-1-5 Kashiwanoha, Kashiwa 277-8583, Japan
        \and
        Zentrum f\"ur Astronomie der Universit\"at Heidelberg, Astronomisches
        Rechen-Institut, M\"onchhofstr. 12-14, D-69120 Heidelberg, Germany
        \and
        Pasadena Consulting Group, 1075 N~Mar Vista Ave, Pasadena, CA 91104 USA
        \and
        Department of Physics and Astronomy, Georgia State University, Atlanta
        GA 30303, USA
        \and
        Zentrum f\"ur Astronomie der Universit\"at Heidelberg, Institut f\"ur
        Theoretische Astrophysik, Philosophenweg 12, D-69120 Heidelberg, Germany
    }

    \date{Received; accepted }

    \abstract{Our ability to predict the structure and evolution of stars is in
       part limited by complex, 3D hydrodynamic processes such as convective
       boundary mixing. Hydrodynamic simulations help us understand the dynamics
       of stellar convection and convective boundaries. However, the codes used
       to compute such simulations are usually tested on extremely simple
       problems and the reliability and reproducibility of their predictions for
       turbulent flows is unclear. We define a test problem involving turbulent
       convection in a plane-parallel box, which leads to mass entrainment from,
       and internal-wave generation in, a stably stratified layer. We compare
       the outputs from the codes \flash, \music, \ppmstar, \prompi, and \slh,
       which have been widely employed to study hydrodynamic problems in stellar
       interiors. The convection is dominated by the largest scales that fit
       into the simulation box. All time-averaged profiles of velocity
       components, fluctuation amplitudes, and fluxes of enthalpy and kinetic
       energy are within $\lesssim 3\sigma$ of the mean of all simulations on a
       given grid ($128^3$ and $256^3$ grid cells), where $\sigma$ describes the
       statistical variation due to the flow's time dependence. They also agree
       well with a $512^3$ reference run. The $128^3$ and $256^3$ simulations
       agree within $9\%$ and $4\%$, respectively, on the total mass entrained
       into the convective layer. The entrainment rate appears to be set by the
       amount of energy that can be converted to work in our setup and details
       of the small-scale flows in the boundary layer seem to be largely
       irrelevant. Our results lend credence to hydrodynamic simulations of
       flows in stellar interiors. We provide in electronic form all outputs of
       our simulations as well as all information needed to reproduce or extend
       our study.}


    \maketitle
%
%-------------------------------------------------------------------

\section{Introduction}

Convection is the most important mixing process in stars. Due to the high
density and opacity deep in the stellar interior, the convection is almost
adiabatic and Mach numbers are typically $\mathcal{M} < 0.05$ in late
evolutionary phases of massive stars and $10^{-4} \lesssim \mathcal{M} \lesssim
10^{-3}$ in early phases, such as core convection during H and He burning. Such
slow flows are nearly incompressible in the sense that ram pressure is much
smaller than thermal pressure, although significant compression and expansion
still occur when fluid packets are displaced radially in the strong, nearly
hydrostatic pressure stratification. The spatial scales involved are so large
that molecular viscosity is negligible and the flow is highly turbulent. The
main effect of convection on the structure and evolution of stars is the
transport of species, energy, and angular momentum. The mixing of chemical
species has important implications for the generation of nuclear energy, origin
of elements, stellar lifetimes, or the observable properties of stars and
stellar populations. The physics of convection is key to determining the
endpoints of stellar evolution which in turn give rise to some of the most
energetic events in the Universe, such as supernova explosions and
compact-object mergers. 

Despite its importance, the treatment of convective mixing and energy transport
is still very rudimentary in stellar-evolution models. This is because the long
thermal and nuclear timescales of stellar evolution make 1D, time-averaged
models the only practical approach. The mixing-length theory
\citep[MLT;][]{prandtl1925a, boehm-vitense:58} is the most common parametric
description of convection in stellar-evolution codes. The MLT's main parameter,
the mixing length $\alpha$, is usually calibrated such that the code reproduces
current properties of the Sun and $\alpha$ is then assumed to be the same in all
convective layers in all stars. Three-dimensional hydrodynamic simulations of
near-surface convection show that $\alpha$ varies across the Hertzsprung-Russel
diagram \citep{trampedach2014b, magic2015, sonoi2019}. \citet{trampedach2014b}
and \citet{jorgensen2018a} illustrate how such realistic 3D models can be used
to replace the MLT in near-surface layers in stellar-evolution calculations.
Deep in the stellar interior, which is the main focus of our work, convection is
so efficient that the stratification becomes isentropic almost independently of
$\alpha$. The MLT is then used to get an estimate of mixing speed, which is
important for the stratification of species when the timescale of some nuclear
reaction(s) becomes comparable to or shorter than the mixing timescale across
the convective layer. Nevertheless, the stellar model remains 1D, and the mixing
is usually described as a diffusive process, although alternative approaches are
being developed \citep[e.g.][]{stephens2021a}.

The MLT is a local theory and as such it cannot describe any mixing caused by
the non-local nature of convection and reaching beyond the formal convective
boundary. Traditionally known as `convective overshooting', this mixing may
involve a number of distinct physical processes and it is better described by
the more general term `convective boundary mixing'
\citep[CBM;][]{denissenkov2013a}. CBM enlarges stellar convection zones and it
is required to explain a large number of observations such as the position of
the turn-off point in open clusters \citep{rosvick1998a}, the width of the main
sequence in the Hertzsprung-Russell diagram \citep{Castro:2014fm}, properties of
double-lined eclipsing binaries \citep[][and references therein]{claret2019a},
or asteroseismic observations of massive stars \citep[][and references
therein]{aerts2021a}. In calculations of stellar evolution, CBM is added in the
form of a simplistic model, such as instantaneous overshooting
\citep{Maeder:1976wg, ekstrom2012a}, a diffusive model \citep{Herwig:2000ua,
battino2016a, baraffe2017lithium}, or an entrainment model
\citep[][]{Staritsin:2013be,Scott:bl}. Each of these formulations has at least
one free parameter, which is usually calibrated by comparison with observations.
However, there is no reason to expect that the same parameter(s) should apply
across different convection zones in different stars at different stages of
evolution.

Multiple research groups have started to use 2D and 3D hydrodynamic codes to
simulate stellar convection and CBM in order to calibrate free parameters of the
CBM models mentioned above, for example, \citet{MeakinArnett2007, jones2017a,
pratt2017extreme, cristini2019a, denissenkov2019a}, and \citet{horst2021a}.
Although the growth rates of convective layers in such simulations seem to be
reasonable for late stages of stellar evolution \citep{MeakinArnett2007,
Woodward:15, jones2017a, mocak2018a}, they are much too high to be compatible
with observations on the main-sequence \citep{MeakinArnett2007, gilet2013a,
Scott:bl}. It is not clear whether this is because the simulations' parameters
are too far from the stellar regime, or it is due to some physics not being
included in the simulations, or it is a sign of numerical problems.

Simulations of convection on the main sequence have become even more attractive
with the recent boom in asteroseismology. The generation of waves by convection
and their propagation through the rest of the star can be observed in 2D and 3D
simulations directly, although most codes require an artificial increase in the
energy generation rate to speed up the flow. The resulting wave spectra can, in
principle, be compared with observations of real stars. However, some groups
obtain spectra with many strong resonance lines \citep{lecoanet2021a} whereas
others get featureless, continuous spectra \citep{alvan2014a, edelmann2019a,
horst2020a}. Again, it is unclear whether the difference can be tracked down to
physical assumptions or they are of numerical origin.

Probably all major hydrodynamics codes used in the field have passed a series of
tests using problems with known solutions such as shock tubes or various
instabilities during their initial growth. However, the stellar hydrodynamic
cases that we are interested in are much more complex and exact solutions do not
exist. Instead, code verification is done by comparing solutions obtained with
different codes. Such exercises have a long tradition in computational dynamics
\citep[][]{Joggerst:da,Ramaprabhu:eu,Dimonte:2004ek} and also in astrophysics
\citep{doherty2010a, beeck2012a, mcnally2012a, kim2014a, kim2016a,
lecoanet2016a, christensen-dalsgaard2020a, silva-aguirre2020a, fleck21a}.
Nonetheless, a test problem focused on the dynamics of convection and mixing
processes in stellar interiors has been missing in the literature.

We constructed a complex test problem as detailed in
Sect.~\ref{sec:test_problem}. It involves stratified, turbulent convection, and
CBM as well as the generation and propagation of internal gravity waves in a
plane-parallel box. The presence of turbulence makes this problem fundamentally
different from classical test cases such as shock tubes. Even if the initial and
boundary conditions are given, the chaotic nature of the flow leads to rapid
amplification of small perturbations in numerical simulations as well as in real
flows. Nevertheless, many space- and time-averaged quantities are expected to
converge upon grid refinement in a statistical way, for instance velocity
profiles, spectra, fluxes, or the cumulative amount of CBM.\footnote{Vorticity
is a good example of a quantity that diverges upon grid refinement in
simulations of inviscid turbulence.} The statistical variation left in these
averages descreases as the length of the averaging time interval is increased.
However, our test problem, designed to be as close as possible to real science
cases, involves continuous heating and mass entrainment. The resulting secular
evolution limits the extent of time averaging applicable and we must estimate
the magnitude of statistical variation when comparing results of such
simulations. Still, we believe that our test problem can be used to compare
codes with accuracy appropriate to the current state of the field.

We ran simulations using the codes \flash, \music, \ppmstar, \prompi, and \slh,
which have been widely used to study hydrodynamic problems in stellar interiors
and are briefly described in Sect.~\ref{sec:codes}. The simulations are compared
in a number of metrics: velocity amplitudes, profiles, and spectra
(Sect.~\ref{sec:velocity_field}), properties of the convective boundary and the
mass entrainment rate (Sect.~\ref{sec:entrainment}), the amplitudes of
fluctuations, and energy fluxes (Sect.~\ref{sec:fluctuations_fluxes}). Finally,
we summarise our results in Sect.~\ref{sec:summary}. Our selection of codes for
this study is based on collaborative ties and availability of resources and it
is far from being unbiased or complete. In particular, our study does not
include any finite-difference and spectral schemes. However, we provide our test
setup as well as all data-analysis scripts in electronic form for anyone
interested in extending this study, see Appendix~\ref{sec:supplement}.

\section{Methods}
\subsection{Motivation of the oxygen shell test problem}

In this study, we aim to define a problem that
involves turbulent convection and CBM while being simple enough to be accessible
to as many codes as possible. We define our test problem to be similar to the
simulations of \citet{jones2017a} and \citet{andrassy2020a} of shell oxygen
burning in a massive star. The stratification of the underlying stellar model is
compared with that of our test problem (specified in
Sect.~\ref{sec:test_problem}) in Fig.~\ref{fig:stratification}. Although the
oxygen shell is spherical, we use plane-parallel geometry both for simplicity
and to reduce computational costs. The lower half of the simulation domain is
initially isentropic and the upper half is stably stratified and approximately
follows the density stratification of the stellar model. Because we intend to
study the dynamics of a single convective boundary in isolation, we do not
include the convective carbon-burning shell, the bottom of which corresponds to
the discontinuity at $r \approx 2.35$ in the stellar model.

To further simplify the problem, we follow \citet{jones2017a} and
\citet{andrassy2020a} and use the ideal-gas equation of state, neglect neutrino
cooling, and replace nuclear reactions with a constant and easy-to-resolve
heating profile. Compared with the original stellar model, the total luminosity
driving convection is $22.5$ times larger and the convective layer contains, due
to its different geometry, $5.4$ times less mass. The resulting rms Mach number
of the convective flow is then ${\approx}\,0.04$, making the problem well
accessible to explicit and implicit codes as well as to codes using low-Mach
approximations. Molecular viscosity, thermal conduction, and radiative
diffusivity are not considered.

We adopt the speed of sound, density, and temperature at the bottom of the
convective layer as units of velocity, density, and temperature, respectively,
and we take the approximate depth of the convective layer to be the unit of
length.\footnote{The convective shell's bottom radius happens to be close to its
radial extent in the stellar model.} The numerical values of these units as well
as those of other units derived from them are summarised in
Table~\ref{tab:units}. The dimensionless problem is specified in detail in the
following section.

\subsection{Problem specification}
\label{sec:test_problem}

\begin{table}
\caption{Basic problem units (upper section), derived units (middle section),
and adopted values of physical constants (lower section).}
\label{tab:units}
\centering
\begin{tabular}{lll}
\toprule
    Quantity & Unit\\
\midrule
    density & $1.820940 \times 10^6$ & g\,cm$^{-3}$\\
    length & $4.000000 \times 10^8$ & cm\\
    temperature & $3.401423 \times 10^9$ & K\\
    velocity & $5.050342 \times 10^8$ & cm\,s$^{-1}$\\
\midrule
    acceleration & $6.376489 \times 10^8$ & cm\,s$^{-2}$\\
    energy & $2.972468 \times 10^{49}$ & erg\\
    luminosity & $3.752995 \times 10^{49}$ & erg\,s$^{-1}$\\
    mass & $1.165402 \times 10^{32}$ & g\\
    pressure & $4.644481 \times 10^{23}$ & dyn\,cm$^{-2}$   \\
    time & $7.920256 \times 10^{-1}$ & s\\
    volume & $6.400000 \times 10^{25}$ & cm$^3$\\
\midrule
    atomic mass unit & $1.660539 \times 10^{-24}$ & g \\
    Boltzmann constant & $1.380649 \times 10^{-16}$ & erg\,K$^{-1}$\\
    gas constant & $8.314463 \times 10^7$ & erg\,g$^{-1}$\,K$^{-1}$\\
    gravitational constant  & $6.674300 \times 10^{-8}$ & dyn\,cm$^2$\,g$^{-2}$\\
\bottomrule
\end{tabular}
\end{table}

\begin{figure}
\includegraphics[width=\linewidth]{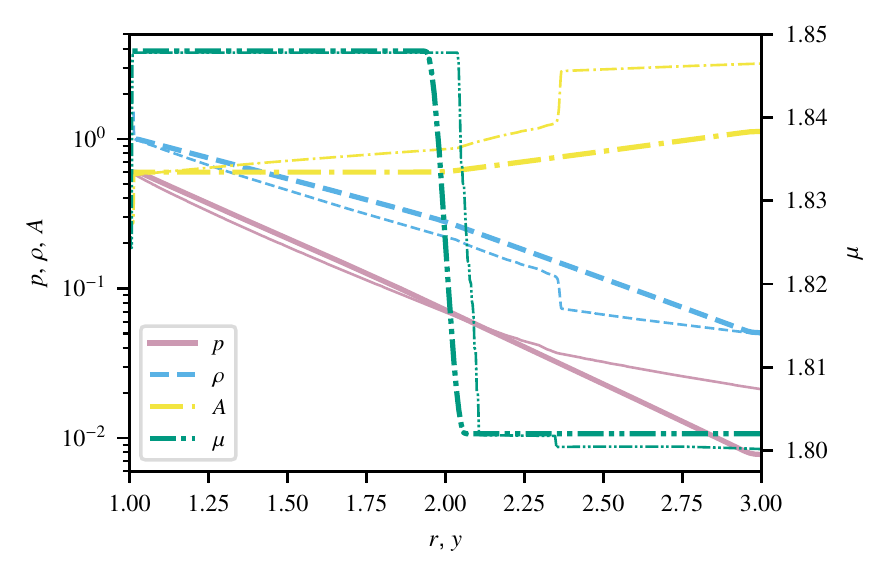}
\caption{Initial stratification in the problem units (Table~\ref{tab:units}) of
   the pressure $p$, density $\rho$, pseudo-entropy $A = p/\rho^\gamma$, and
   mean molecular weight $\mu$ in the stellar model (thin lines) and in the test
   problem (thick lines). The discontinuity visible in the stellar model at $r
   \approx 2.35$ is the bottom of the carbon-burning shell, which is not
   included in our test problem.}
\label{fig:stratification}
\end{figure}

The initial hydrostatic stratification, shown in Fig.~\ref{fig:stratification},
is computed using the following profile of gravitational acceleration:
\begin{equation}
    g(y) = g_0\,f_g(y)\,y^{-5/4},
    \label{eq:gravity}
\end{equation}
where $g_0 = 1.414870$ is the gravitational acceleration at the bottom of the
oxygen-burning shell in the stellar model. The vertical coordinate $y$ runs from
$1$ to $3$, which in principle allows the same stratification to be used in
spherical coordinates simply by setting the radius $r = y$. This profile is
similar to the stellar one and the reduction in gravity with height helps
prevent a too-fast decrease in the pressure scale height. We `turn off'
gravity close to the lower and upper boundaries of the simulation domain using
the factor
\begin{equation}
    f_g(y) = 
    \begin{cases}
        \frac{1}{2}\left\{1 + \sin\left[ 16\pi\left(y -\frac{1}{32}\right)
        \right]\right\},& \mathrm{for}\ 1 \le y < 1 + \frac{1}{16}, \\[0.5em]
        1,& \mathrm{for}\ 1 + \frac{1}{16} \le y \le 3 - \frac{1}{16}, \\[0.5em]
        \frac{1}{2}\left\{1 + \sin\left[ 16\pi\left(y -\frac{1}{32}\right)
        \right]\right\},& \mathrm{for}\ 3 - \frac{1}{16} < y \le 3,
    \end{cases}
    \label{eq:fg}
\end{equation}
which forces density and pressure to become constant close to the boundaries and
thus makes possible the use of a simple reflective boundary condition there. The
computational domain contains a mixture of two monatomic ideal gases with
\mbox{$\gamma = 5/3$} and mean molecular weights of \mbox{$\mu_0 = 1.848$} and
$\mu_1 = 1.802$. Initially, the convective layer is filled with the $\mu_0$
fluid and the stable layer with the $\mu_1$ fluid. There is a smooth transition
between the two layers and the fractional volume $\eta_1$ of the $\mu_1$ fluid
varies as
\begin{equation}
    \eta_1(y) = 
    \begin{cases}
        0,& \mathrm{for}\ 1 \le y < 2 - \frac{1}{16}, \\[0.5em]
        \frac{1}{2}\left[1 + \sin\left( 8\pi y \right)\right],& \mathrm{for}\ 2
        - \frac{1}{16} \le y \le 2 + \frac{1}{16}, \\[0.5em]
        1,& \mathrm{for}\ 2 + \frac{1}{16} < y < 3.
    \end{cases}
    \label{eq:eta1}
\end{equation}
The mixture is assumed to be in local pressure and thermal equilibrium
everywhere and the stratification follows the piecewise-polytropic
pressure-density relation
\begin{equation}
    \frac{\mathrm{d}\ln p}{\mathrm{d}\ln \rho} = 
    \begin{cases}
        \gamma_0,& \mathrm{for}\ 1 \le y < 2 - \frac{1}{16}, \\[0.5em]
        \gamma_0 + \eta_1(y)(\gamma_1 - \gamma_0),&
        \mathrm{for}\ 2 - \frac{1}{16} \le y \le 2 + \frac{1}{16}, \\[0.5em]
        \gamma_1,& \mathrm{for}\ 2 + \frac{1}{16} < y < 3,
    \end{cases}
    \label{eq:dlnp_by_dlnrho}
\end{equation}
where $\gamma_0 = \nicefrac{5}{3}$ and $\gamma_1 = 1.3$.
Equations~\ref{eq:gravity}\,--\,\ref{eq:dlnp_by_dlnrho}, together with the
ideal-gas law define a unique hydrostatic state.

We impose friction-free, non-conductive wall boundary conditions at $y = 1$ and
$y = 3$. The specific implementation of these boundary conditions is
code-dependent, see Sect.~\ref{sec:codes}. The computational domain is periodic
in the two horizontal dimensions and spans the coordinate intervals \mbox{$-1
\le x \le 1$} and $-1 \le z \le 1$. The initial aspect ratio of the convective
layer is thus $2{:}1$ (width to height).

Convection is driven using a time-independent heat source concentrated close to
the lower boundary of the convective layer with energy generation rate per unit
volume of
\begin{equation}
    \dot{q}(y) =
    \begin{cases}
       \dot{q}_0 \sin\left( 8\pi y \right),& \mathrm{for}\ 1 \le y \le 1 +
       \frac{1}{8}, \\[0.5em]
        0,& \mathrm{for}\ 1 + \frac{1}{8} < y \le 3,
    \end{cases}
\end{equation}
where $\dot{q}_0 = 3.795720 \times 10^{-4}$. The total luminosity of the heat
source is $1.2082151 \times 10^{-4}$. When the problem is discretised, we take
into account the fact that the average value of $\dot{q}(y)$ over a
computational cell of height $\Delta y$ centred around $y = y_0$ is
$\dot{q}(y_0) \sin(4\pi \Delta y) / (4\pi \Delta y)$. The heat source is defined
to be smooth and easy to resolve for this study; the energy generation profile
in the stellar model is asymmetric and discontinuous \citep[see Fig.~3
of\ ][]{jones2017a}. We do not include any cooling term. The Kelvin-Helmholtz
timescales of the convective layer and of the whole computational domain are
$1.29 \times 10^4$ and $1.43 \times 10^4$ time units, respectively.

To break the initial symmetry, we use a density perturbation
\begin{equation}
    \frac{\Delta \rho}{\rho_0} = 5 \times 10^{-5}\,\frac{\dot{q}(y)}{\dot{q}_0}
    \left[ \sin(3\pi x) + \cos(\pi x) \right]
    \left[ \sin(3\pi z) - \cos(\pi z) \right],
\label{eq:perturbation}
\end{equation}
where $\rho_0 = 1$ is the density at the bottom of the convective layer. The
perturbation only affects the heating layer, so it is also concentrated close
to the lower boundary of the convective layer. There is no pressure
perturbation. The smooth density perturbation, although small, allows us to
produce a well-resolved initial transient, which should be similar in all codes.

The problem is described by the inviscid Euler equations with gravity and volume
heating,
\begin{align}
   \frac{\partial \rho}{\partial t} + \bm{\nabla} \cdot (\rho \bm{V}) &= 0,
   \label{eq:mass_conservation} \\
   \frac{\partial (\rho \bm{V})}{\partial t} + \bm{\nabla} \cdot (\rho \bm{V}
   \otimes \bm{V} + p \bm{I}) &= \rho \bm{g}, \label{eq:momentum_conservation}
   \\
   \frac{\partial (\rho E)}{\partial t} + \bm{\nabla} \cdot [( \rho E + p )
   \bm{V} ] &= \rho \bm{V} \cdot \bm{g} + \dot{q},
   \label{eq:energy_conservation_1}
\end{align}
where $\bm{V}$ is the velocity vector, $\bm{I}$ the unit tensor, $\bm{g}$ the
gravitational acceleration vector pointed towards the negative $y$ axis, and $E
= e + \frac{1}{2} \|\bm{V}\|^2$  the specific total energy, which includes the
specific internal energy $e$ and the specific kinetic energy $\frac{1}{2}
\|\bm{V}\|^2$. Some of our codes, as indicated in Table~\ref{tab:codes}, include
the specific potential energy $\Phi$ in the total energy $E_\Phi = E + \Phi$
and, instead of evolving Eq.~\ref{eq:energy_conservation_1}, they evolve the
equivalent equation
\begin{equation}
   \frac{\partial (\rho E_\Phi)}{\partial t} + \bm{\nabla} \cdot [( \rho E_\Phi
   + p ) \bm{V} ] =  \dot{q}.
   \label{eq:energy_conservation_2}
\end{equation}
The system of equations is closed by the ideal-gas law
\begin{equation}
   p = (\gamma - 1)\, \rho e
\end{equation}
with $\gamma = \nicefrac{5}{3}$. We track the mixing between the two layers by
advecting the partial density $\rho X_i$ of either of the two fluids,
\begin{equation}
   \frac{\partial (\rho X_i)}{\partial t} + \bm{\nabla} \cdot (\rho X_i \bm{V})
   = 0, \label{eq:composition} \\
\end{equation}
and the other mass fraction follows from the requirement \mbox{$X_0 + X_1 = 1$}.
The mass fraction $X_i$ acts as a passive tracer of mixing in our setup, which
is a consequence of the set of assumptions we make. The passive nature of
composition may be somewhat surprising since composition has a direct influence
on buoyancy. However, fluctuations in composition are in our setup advected
together with fluctuations in entropy and the latter determines the buoyancy. It
does not matter whether the entropy difference between a fluid parcel and its
surroundings is a result of a difference in composition or heat content or
both. This special property of our setup would be lost if we introduced a
composition dependence in
Eqs.~\ref{eq:mass_conservation}--\ref{eq:energy_conservation_2} by using a more
complex equation of state, or by including temperature- and
composition-dependent terms such as heat-conduction, radiative-diffusion, or
nuclear reactions.

\subsection{Codes}
\label{sec:codes}

We run simulations of the problem described above using \numcodes\ 3D
hydrodynamic codes that are well established in the field of stellar convection:
\flash, \music, \ppmstar, \prompi, and \slh. All of these codes are based on the
finite-volume method but there are many differences between the numerical
schemes as shown in Table~\ref{tab:codes} and in the following subsections.

\begin{table*}
\caption{Basic characteristics of the codes used in this study.}
\label{tab:codes}
\centering
\begin{tabular}{llllllll}
\toprule
   Code & Energy   & Grid & Reconstruction & Numerical Flux & Dimensional & Time    & FP \\
        & Equation &      &                & Function       & Splitting   & Stepper & Precision \\
\midrule
   \flash & Eq.~\ref{eq:energy_conservation_1} & collocated & PPM & HLLC
          & no & \citet{lee2009a} & $64$-bit$^*$ \\
        &  &  & & \citet{toro2009a} & & & \\
   \music & Eq.~\ref{eq:energy_conservation_1} & staggered & \citet{van1974towards}
          & upwinded advection & no & Crank-Nicolson & $64$-bit \\
        &  &  & & \citet{viallet2016jacobian} & & & \\
   \ppmstar & Eq.~\ref{eq:energy_conservation_1} & collocated & PPM+PPB &
            \citet{woodward07} & yes & PPM$^{**}$ & $32$-bit \\
   \prompi & Eq.~\ref{eq:energy_conservation_1} & collocated & PPM &
           \citet{Fryxell2000} & yes & \citet{Fryxell2000} & $64$-bit$^*$ \\
   \slh & Eq.~\ref{eq:energy_conservation_2} & collocated & PPM & AUSM+-up & no
        & RK3 & $64$-bit \\
        &  &  & & \citet{liou2006a} & & & \\
\bottomrule
\end{tabular}
\vspace{0.5em}
\tablefoot{Sect.~\ref{sec:codes} contains further references and details.
$^*$Output is written in $32$-bit precision. $^{**}$Exact algorithm to be
described in a future publication.}
\end{table*}

\subsubsection{\flash}

\flash\ \citep{Fryxell2000} is a modular multidimensional hydrodynamics code
that originated from the combination of the legacy PROMETHEUS code
\citep{Fryxell1989} and the AMR library PARAMESH \citep{MacNeice2000}. \flash\
was originally developed to simulate Type Ia supernovae
\citep[e.g.][]{plewa2004} and has since been extended by a large variety of
modules including magnetic fields, radiation transfer, and the consideration of
a cosmological redshift. Due to its great flexibility, \flash\ has since been
used to address various astrophysical problems including core-collapse
\citep[e.g.][]{Couch2015} and type Ia \citep[e.g.][]{willcox2016a} supernova
explosions, galaxy evolution \citep[e.g][]{Scannapieco2015}, or interstellar
turbulence \citep[e.g][]{Federrath2010}.  For this work we used version 4.6.2 of
\flash\ with its default unsplit hydrodynamics solver. Compared to the default
settings, we increase the order of reconstruction to the 3rd order PPM
\citep{ColellaWoodward1984} and apply the HLLC Riemann solver. The top and
bottom boundary are implemented as reflective boundaries. \flash\ uses double
precision (64-bit) floating-point arithmetic to perform the computations, but
the output that is used for post-processing has been written in single precision
(32-bit) to save disk space.  

\subsubsection{\music}

The MUlti-dimensional Stellar Implicit Code \music\
\citep{viallet2016jacobian,goffrey2017benchmarking} is a time-implicit
hydrodynamics code designed to study key phases of stellar evolution in two and
three dimensions in spherical or Cartesian coordinates. The code solves the
Euler equations, optionally supplemented with diffusive radiation transport,
gravity, the Coriolis and centrifugal terms, and active and/or passive scalars.
The equation set is closed using either an ideal gas or a tabulated equation of
state. The equations are spatially discretised using a finite volume method, on
a staggered mesh, with scalar quantities defined at cell centres, and vector
components at cell faces. The advection step uses a second-order interpolation,
and a gradient limiter originally described by van Leer \citep{van1974towards}.
Time discretisation is carried out using the Crank-Nicolson method, and a
physics-based preconditioner is used to accelerate the convergence of the
implicit method \citep{viallet2016jacobian}. The boundary conditions are
implemented via appropriate ghost zone layers (reflective along the top and
bottom walls and periodic along the side walls), and the code uses 64-bit
precision throughout. 

The code has been benchmarked against a number of standard hydrodynamical test
problems \citep[e.g.][]{goffrey2017benchmarking} and has been applied to a
number of stellar physics problems, including accretion \citep{geroux2016multi}
and convective overshooting \citep{pratt2017extreme,baraffe2017lithium,pratt2020}.

The concentration of the $\mu_1$ fluid is advected as a passive scalar. The
corresponding flux is reconstructed using the mass fractions of the $\mu_1$
fluid. For comparison with the other codes, all output is linearly interpolated
onto a cell-centred grid as a first post-processing step.
 
\subsubsection{\ppmstar}

The explicit Cartesian compressible gas-dynamics code \ppmstar\ is based on the
Piecewise-Parabolic Method \citep[PPM;][]{woodward_colella81,
woodward_colella84,ColellaWoodward1984,woodward:86,woodward07}. In its most
recent version \citep{Woodward:2019ip} it solves the conservation equations in a
perturbation formulation with regard to an initial base state that is valid for
perturbations of any size. This allows the computation to be carried out with
only 32-bit precision and roughly doubles the execution speed. The time-stepping
algorithm has been revised and will be described in a future publication.
Another key feature of \ppmstar\ is tracking the advection of the concentrations
in a two-fluid scheme using the Piecewise-Parabolic Boltzmann moment-conserving
advection scheme \citep[PPB;][]{woodward:86,Woodward:15}. Nuclear reactions and
energy production is taken into account with approximate networks
\citep{herwig:14,andrassy2020a,stephens2021a}. Both radiation pressure and
diffusion can be taken into account (Mao et al.\ in prep.). Reflective boundary
conditions are used at the top and bottom boundaries.

\subsubsection{\prompi}

\prompi\ is a multidimensional hydrodynamics code based on an Eulerian
implementation of the piecewise parabolic method PPM by
\citet{ColellaWoodward1984} capable of treating realistic equations of state
\citep{Colella1985} and multi-species advection. It is equipped with an equation
of state to handle the semi-degenerate stellar plasma \citep{timmes2000a},
gravity, radiative diffusion and a general nuclear reaction network. \prompi\ is
a version of the legacy \texttt{PROMETHEUS} code \citep{Fryxell1991}
parallelised with MPI (Message-Passing-Interface) by \citet{MeakinArnett2007}.
Notable scientific work enabled by \prompi\ can by found in
\citet{MeakinArnett2007,ArnettMeakin2009,VialletMeakin2013,mocak2018a}, or
\citet{cristini2019a}. Latest development of \prompi\ includes GPU accelleration
(Hirschi, private communication) and runtime calculation of space-time averaged
mean-fields for extensive Reynolds-Averaged-Navier Stokes (or RANS)
analysis\footnote{For more details on the \prompi's mean-fields utilisation see
the ransX framework
\href{https://github.com/mmicromegas/ransX}{https://github.com/mmicromegas/ransX}}.
\prompi\ uses $64$-bit precision internally but it writes output in $32$-bit
precision to save disk space. Reflective boundary conditions are used at the top
and bottom boundaries.

\subsubsection{\slh}
\label{sec:slh}

The Seven-League Hydro (\slh) code, initially developed by \citet{miczek2013a},
solves the fully compressible Euler equations in one to three spatial
dimensions. It contains a general equation of state including radiation pressure
and electron degeneracy \citep{timmes2000a} and supports radiative transfer in
the diffusion limit. A monopole and a full 3D gravity solvers are also
available. An arbitrary number of fluids can be advected and interactions
between them can be simulated using a nuclear-reaction module
\citep{edelmann2014a}.

The equations are discretised on logically rectangular, but otherwise arbitrary,
curvilinear grids using a finite-volume scheme. The code specialises in slow,
nearly hydrostatic flows in the stellar interior. Various methods to treat the
hydrostatic background stratification \citep{edelmann2021a} are used in
combination with several low-Mach flux functions \citep{liou2006a, li2008a,
miczek2015a} to reduce dissipation at low Mach numbers, which is unacceptably
high with standard flux functions. Reconstruction schemes available range from
constant through linear with several optional slope limiters and unlimited
parabolic to the PPM reconstruction of \citet{ColellaWoodward1984}. \slh\
supports both implicit and explicit time stepping. The code has been shown to
scale up to several hundred thousand cores \citep{edelmann2016b, hammer2016a}
and applied to problems involving mixing processes
\citep{edelmann2017a,horst2021a} and wave generation \citep{horst2020a} in
stellar interiors.

In the present work, we use PPM reconstruction with a slightly modified version
of the AUSM+-up flux function \citep{liou2006a} and the Deviation well-balancing
method of \citet{berberich2021a}. The wall boundary conditions are implemented
as flux-based boundaries such that mass and energy fluxes through the walls are
exactly zero. Ghost cells are used at the wall boundaries in the reconstruction
process: we perform parabolic extrapolation for all conserved variables with the
exception of composition variables, which are assumed to be constant at the wall
boundaries. Because the flow in the test problem is relatively fast, we employ
explicit time stepping with the RK3 scheme to reduce computational costs. The
code uses 64-bit floating-point arithmetic. Unlike the other codes, we add
another density perturbation on top of that defined by Eq.~\ref{eq:perturbation}
in the form of white noise with an amplitude of $5 \times 10^{-7}$ because \slh\
otherwise preserves the reflection symmetry of Eq.~\ref{eq:perturbation} with
respect to the plane $x = -z$.

\subsection{Simulations and their output}

We use the same Cartesian computational grids with constant spacing in all
codes: a low-resolution grid with $128^3$ cells, a medium-resolution grid of
$256^3$ cells and a high-resolution grid of $512^3$ cells. Because all
quantities we compare in Sect.~\ref{sec:results} converge rapidly upon grid
refinement, we only perform a full $512^3$ run with \ppmstar\ and a short one
with \prompi\ to save computing time.

All simulations are stopped at time $t_\mathrm{end} = 2 \times 10^3$, which
corresponds to $25$ convective turnover timescales, and we write output every
$5$ time units with the exception of the $512^3$ \prompi\ run, in which the
output is written every $1.266$ time units. The output includes the full 3D
state information, which is post-processed to obtain 1D horizontal averages for
a number of quantities, kinetic-energy spectra, and 2D slices through the
simulation box, see Appendix~\ref{sec:supplement} for details. We use both
horizontal volume-weighted averages
\begin{equation}
   \vw{q}_j = \frac{1}{N_x N_z} \sum_{i,k} q_{i,j,k},
   \label{eq:vw_average}
\end{equation}
and horizontal mass-weighted averages
\begin{equation}
   \mw{q}_j = \frac{\sum_{i,k} q_{i,j,k}\ \rho_{i,j,k}}{\sum_{i,k}
   \rho_{i,j,k}},
   \label{eq:mw_average}
\end{equation}
where $q$ is the quantity to be averaged, $\rho$ is the density, $i$, $j$, and
$k$, are grid cell indices along the $x$, $y$, and $z$ axes, respectively, and
$N_x$ and $N_z$ are the total numbers of cells along the axis given in the
subscript. The cell volume does not appear in Eqs.~\ref{eq:vw_average} and
\ref{eq:mw_average}, because it is the same for all cells. We use the notation
$\langle q \rangle$ for time averages. We always give the averaging time
interval in the text or in the figure caption. In some cases, we need to smooth
a time series $q(t)$ to suppress noise and make it easier to visually compare
different simulations. To do so, a centred top-hat convolution filter is
employed. Its width $\tauav$ is specified in each case individually. We suppress
boundary effects by padding the time series with the time average of $q(t)$
during the first and last $\frac{\tauav}{2}$ time units before performing the
convolution.

\section{Results}
\label{sec:results}

\begin{figure*}[h!]
  \centering
  \includegraphics[width=0.95\textwidth]{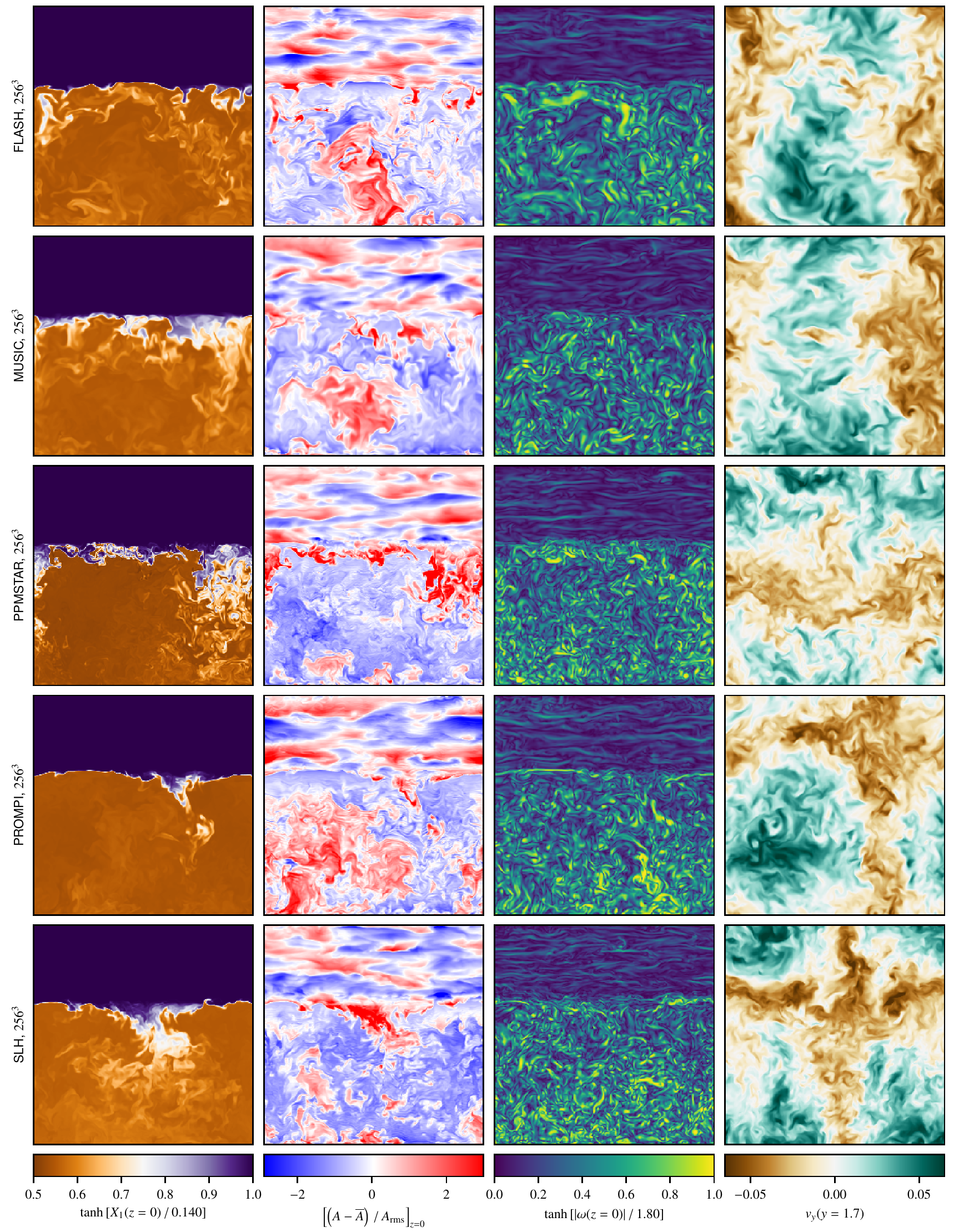}
  \caption{Renderings of the flow field at $t \,{=}\, 1000$. The variables shown
     (from left to right) are the mass fraction $X_1$, relative entropy
     fluctuations $(A - \overline{A}) / A_\mathrm{rms}$, magnitude of vorticity
     $|\omega|$, and the vertical component of velocity $v_y$. The first three
     variables correspond to slices through the simulations box in the $z
     \,{=}\, 0$ plane and the last one in the $y \,{=}\, 1.7$ plane. The
     non-linear scaling of $X_1$ and $|\omega|$ and the normalisation of entropy
     fluctuations by their horizontal rms spread $A_\mathrm{rms}$ (see the
     colour bars) are introduced for visualisation purposes.}
\label{fig:slices}
\end{figure*}

Owing to the well-resolved initial perturbation, the initial growth of
convection proceeds at the same rate in all of the codes considered (see
Fig.~\ref{fig:velocity_evolution} and Sect.~\ref{sec:velocity_field}). As the
animations available on
Zenodo\footnote{\url{https://doi.org/10.5281/zenodo.5796842}} show, a
substantial amount of small-scale structure appears in the large-scale hot
bubbles during their initial rise from the heating layer towards the top of the
convective layer. They deform the convective-stable interface significantly when
they impact it at $t\,{\approx}\,60$, generating the first
upward-propagating internal gravity waves (IGW). The stable nature of the upper
layer forces the convective upflows to decelerate and, ultimately, reverse. As
this happens, the flows drag some of the $\mu_1$ fluid into the convective
layer, starting the process of mass entrainment, see Fig.~\ref{fig:slices}. The
flow keeps accelerating during the first few convective timescales $\tauconv =
80$ (see Eq.~\ref{eq:tauconv}) and we conservatively define the end of this
initial transient to be $t_0 = 500 \approx 6\,\tauconv$. We focus our analysis
on the remaining $1500$ time units (${\approx}19\,\tauconv$) of steady
convection accompanied with continuous increase in the convective layer's mass
due to mass entrainment. In the following subsections, we present different
aspects of the simulations and compare their evolution in the \numcodes{} codes
in detail.

\subsection{Velocity field}
\label{sec:velocity_field}

\begin{figure*}
\begin{minipage}{.5\textwidth}
  \centering
  \includegraphics[width=\textwidth]{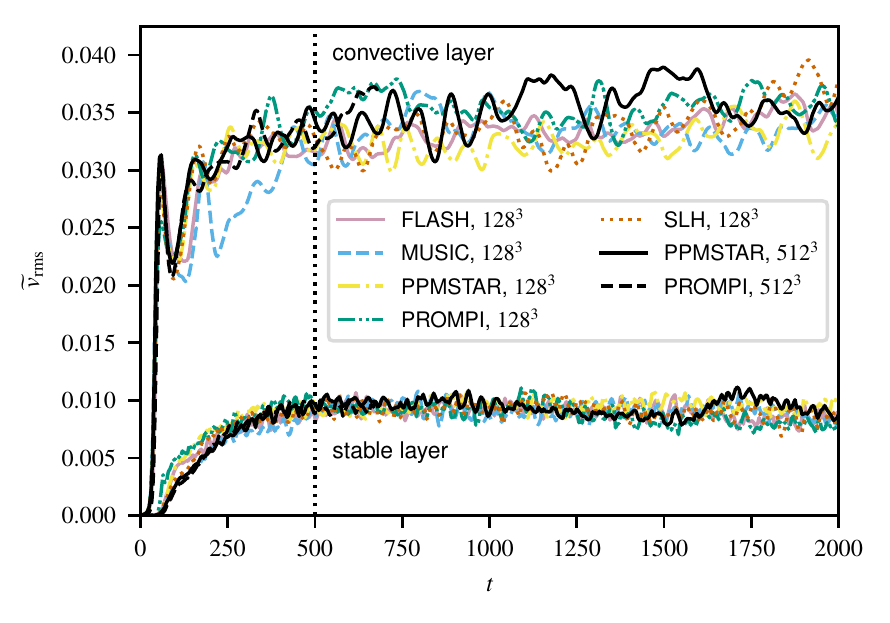}
\end{minipage}%
\begin{minipage}{.5\textwidth}
  \centering
  \includegraphics[width=\textwidth]{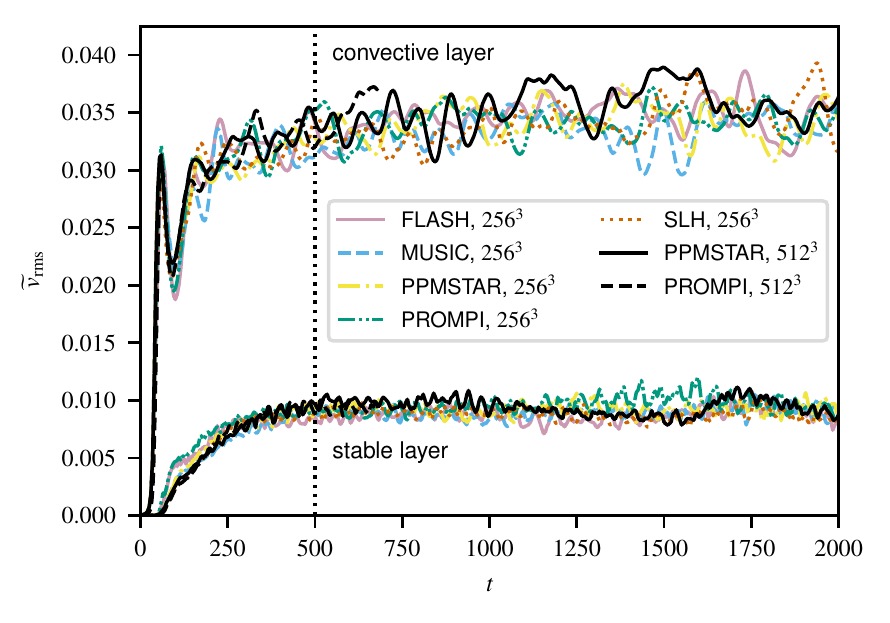}
\end{minipage}
\caption{Time evolution of the mass-weighted average rms velocity fluctuations
   in the convection zone ($y < y_\mathrm{ub}(t) - 0.1$, upper set of curves)
   and in the stable layer ($y > y_\mathrm{ub}(t) + 0.1$, lower set of curves).
   The location $y_\mathrm{ub}(t)$ of the convective boundary is tracked in time
   as shown in Fig.~\ref{fig:yub}. The dotted vertical line marks the end of the
   initial transient excluded from the analysis.}
\label{fig:velocity_evolution}
\end{figure*}

We first compare the simulations in terms of horizontally averaged rms velocity
fluctuations
\begin{equation}
   \mw{v}_\mathrm{rms}(y) = \left( \mw{\sigma}_{v,x}^{\,2} +
   \mw{\sigma}_{v,y}^{\,2} + \mw{\sigma}_{v,z}^{\,2} \right) ^\frac{1}{2},
\end{equation}
where $\mw{\sigma}_{v,x}$, $\mw{\sigma}_{v,y}$, and $\mw{\sigma}_{v,z}$ are
standard deviations of the three components of the velocity vector. We further
average the velocity profiles over the convective and stable layer to obtain
(mass-weighted) bulk measures of typical velocity fluctuations in the two
layers.\footnote{We do not use any special notation for the bulk averages. It
should be clear from the text whether a bulk quantity or the vertical dependence
$\mw{v}_\mathrm{rms}(y)$ is being discussed.} Because a substantial amount of
mass gets entrained into the convective layer during the simulations, we track
the vertical coordinate $y_\mathrm{ub}$ of the upper boundary of the convective
layer in time as described in Sect.~\ref{sec:entrainment}. We compute the
averages in the regions with $y < y_\mathrm{ub}(t) - 0.1$ (convective layer) and
$y > y_\mathrm{ub}(t) + 0.1$ (stable layer). The offsets of $0.1$ length units
are used to exclude the transition zone.

The time evolution of the rms velocity fluctuations in the convective and stable
layer is shown in Fig.~\ref{fig:velocity_evolution}. During the initial
transient ($t < t_0 = 500$), the rms velocity in the convective layer increases
until it has saturated at a mean value of ${\approx}\,0.034$. Because the speed
of sound is approximately $0.8$ in the middle of the convective layer, the
typical Mach number is ${\approx}\,0.04$. The flow speed is statistically the
same in all of our simulations. This means that even in the $128^3$ simulations
the convection is a fully developed turbulent flow, in which kinetic energy is
dissipated at a rate independent of the code-dependent numerical viscosity close
to the grid scale. The $3\sigma$ fluctuations around the mean value of
$\mw{v}_\mathrm{rms}$ range from $11\%$ to $19\%$ with a median value of $13\%$.
The $512^3$ \ppmstar\ run contains two high-velocity episodes in the time
intervals $1090 \,{\lesssim}\, t \,{\lesssim}\, 1250$ and $1440 \,{\lesssim}\, t
\,{\lesssim}\, 1620$, both ${\approx}\,2\tauconv$ long, during which the bulk
convective velocity increases by up to ${\approx}\,15\%$ as compared with the
$128^3$ and $256^3$ runs. These episodes are likely of statistical origin,
although we would need an ensemble of $512^3$ runs to confirm this hypothesis.
When all of the simulations are considered, there is some weak evidence for a
slight systematic increase in velocity in the interval $500 < t < 2000$ with a
median value of $4\%$. However, run-to-run values range from $-1\%$ to $15\%$
and seem to be dominated by statistical variation, so we do not subtract the
linear trend when computing the magnitude of fluctuations.

The rms velocity in the stable layer initially increases slowly as IGW generated
at the convective-stable interface propagate through the stable layer. To
understand the slow vertical propagation of the waves, we refer to the 2D
renderings of entropy fluctuations in Fig.~\ref{fig:slices}. The renderings
show a number of wavefronts spanning almost the full width of the computational
domain and inclined at angles of $5^\circ \lesssim \vartheta \lesssim 15^\circ$
with respect to the horizontal, ignoring projection effects. The
Brunt-V\"ais\"al\"a frequency $N_\mathrm{BV}$ does not change much across the
stable layer and its typical value is $N_\mathrm{BV} = 0.42$ at $t = 0$. Using
linear theory of IGW \citep[see e.g.][]{sutherland2010}, we estimate that such
long waves have periods in the range $2 \tauconv \gtrsim P_{\mathrm{IGW}}
\gtrsim 0.7 \tauconv$. The magnitude of the vertical component of their group
velocity is $0.001 \lesssim v_{y,\mathrm{IGW}} \lesssim 0.01$, implying that it
takes the waves between ${\approx}100$ and ${\approx}1000$ time units to
propagate from the initial convective-stable interface at $y = 2$ to the upper
boundary condition at $y = 3$. Projection effects, which decrease the apparent
angle $\vartheta$ in the 2D slices, make this estimate slightly biased towards
longer timescales. Nevertheless, Fig.~\ref{fig:velocity_evolution} shows that
the final velocity amplitude of ${\approx}\,0.009$ in the stable layer is
reached by $t \,{=}\, t_0 \,{=}\, 500$ in all of our simulations. This
observation, combined with the estimate above suggests that there cannot be
strong resonant wave amplification caused by many wave reflections between the
upper boundary of the simulation box and the convective-stable interface because
that would cause the rms velocity to increase on much longer timescales.
Finally, the fact that the rms velocity reaches the same value on grids ranging
from $128^3$ to $512^3$ implies that the dominant wave patterns are well
resolved even on the coarsest grid. This conclusion is further supported by
kinetic-energy spectra, which we discuss at the end of this section.

\begin{figure}
\includegraphics[width=\linewidth]{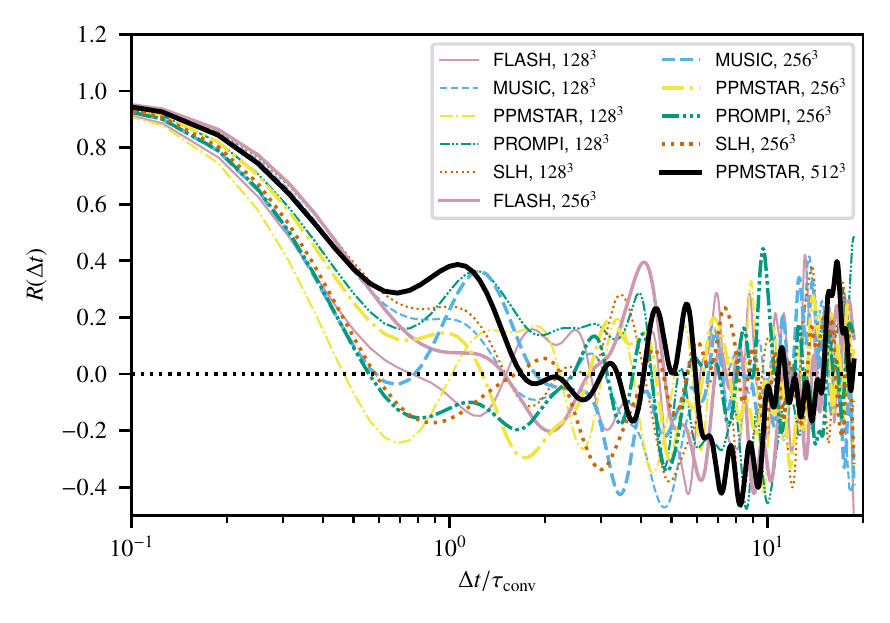}
\caption{Autocorrelation function $R(\Delta t)$ of the convective velocity as a
function of the time shift $\Delta t$ in units of the convective turnover
timescale $\tauconv$. See Eq.~\ref{eq:autocorrelation} and the associated text
in detail.}
\label{fig:autocorrelation}
\end{figure}

We define the convective turnover timescale to be
\begin{equation}
   \tauconv = \frac{2 \langle \Delta y_\mathrm{cl} \rangle}{\langle
   \widetilde{v}_\mathrm{rms} \rangle} = 80,
   \label{eq:tauconv}
\end{equation}
where $\langle \widetilde{v}_\mathrm{rms} \rangle = 0.034$ is the
above-mentioned characteristic convective velocity and
\begin{equation}
   \langle \Delta y_\mathrm{cl} \rangle = \langle y_\mathrm{ub} \rangle -
   y_\mathrm{bot} = 1.36
\end{equation}
is the average depth of the convective layer, that is the distance between the
layer's bottom at $y_\mathrm{bot} = 1$ and the vertical coordinate
$y_\mathrm{ub}$ of the upper convective boundary averaged in the time interval
$500 \le t \le 2000$, see Fig.~\ref{fig:yub} and Sect.~\ref{sec:entrainment} for
details. To show how well this timescale describes variability in the global
convective velocity field, we compute the autocorrelation function
\begin{equation}
   R(\Delta t) = \frac{\int_{t_0}^{t_\mathrm{end}}
   \widetilde{v}_\mathrm{rms}^{\,*}(t)\ \widetilde{v}_\mathrm{rms}^{\,*}(t +
\Delta t)\, \mathrm{d}t}{\int_{t_0}^{t_\mathrm{end}}
\widetilde{v}_\mathrm{rms}^{\,*}(t)^2\, \mathrm{d}t},
\label{eq:autocorrelation}
\end{equation}
where $\Delta t$ is a time shift and $\widetilde{v}_\mathrm{rms}^{\,*}(t)$ is
constructed from the bulk convective velocity $\widetilde{v}_\mathrm{rms}(t)$ as
follows: (1) the initial transient ($t < t_0$) is discarded, (2) the best-fit
linear trend is subtracted to suppress spurious correlations caused by any
slight systematic changes in $\widetilde{v}_\mathrm{rms}(t)$ on long timescales,
and (3) the resulting time series is made periodic by appending to it a
time-reversed version of itself. Figure~\ref{fig:autocorrelation} shows that
$R(\Delta t)$ reaches high values for $\Delta t \lesssim 0.5 \tauconv$, implying
that the time series is strongly correlated on such short timescales. However,
$R(\Delta t)$ decreases steeply and, although the function's first zero crossing
occurs at $\Delta t < \tauconv$ in some runs and at $\Delta t > \tauconv$ in
others due to stochasticity, $\Delta t \approx \tauconv$ is generally a good
estimate of the temporal spacing between different, largely independent, flow
realisations.

\begin{figure*}
\begin{minipage}{.5\textwidth}
  \centering
  \includegraphics[width=\textwidth]{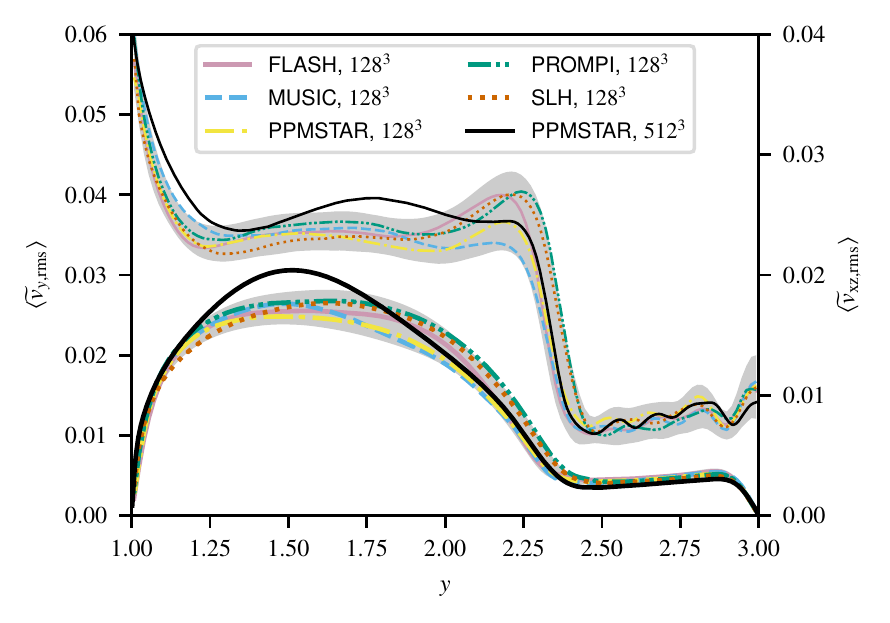}
\end{minipage}%
\begin{minipage}{.5\textwidth}
  \centering
  \includegraphics[width=\textwidth]{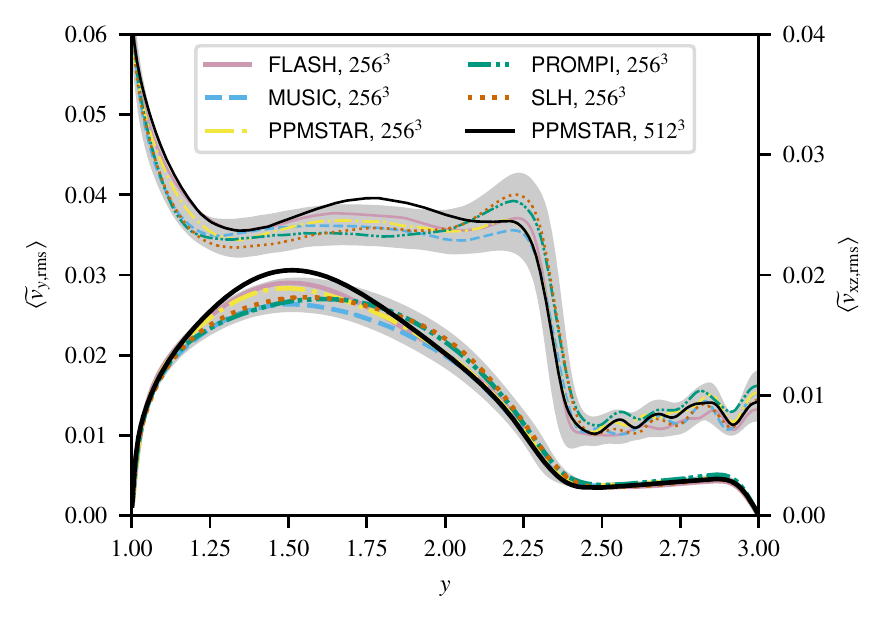}
\end{minipage}
\caption{Vertical profiles of the rms velocity components $\langle
   \widetilde{v}_{y,\mathrm{rms}} \rangle$ (thick lines) and $\langle
   \widetilde{v}_\mathrm{xz,rms} \rangle$ (thin lines) in the vertical and
   horizontal directions, respectively, averaged in a time interval $\tauav =
   6\tauconv = 480$ wide and centred on time $1250$. The two velocity axes in
   each plot have different scales to avoid the curves' overlapping. The grey
   bands give an estimate of $\pm 3 \sigma$ statistical variation in the
   averages due to stochasticity, see Sect.~\ref{sec:velocity_field} for
   details.}
\label{fig:velocity_profiles}
\end{figure*}

Figure~\ref{fig:velocity_profiles} shows time-averaged profiles of rms vertical and
horizontal velocity fluctuations,
\begin{align}
   \mw{v}_{y,\mathrm{rms}}(y) &= \mw{\sigma}_{v,y}, \\
   \mw{v}_{xz,\mathrm{rms}}(y) &= \left( \mw{\sigma}_{v,x}^{\,2} +
   \mw{\sigma}_{v,z}^{\,2} \right) ^\frac{1}{2}.
\end{align}
The vertical fluctuations $\langle \widetilde{v}_{y,\mathrm{rms}} \rangle$ reach
a broad maximum close to the middle of the convective layer. They vanish at the
bottom of the simulations box as required by the wall boundary conditions but
they also drop by as much as a factor of four at the transition to the stable
layer. The velocity field is dominated by the horizontal component $\langle
\widetilde{v}_\mathrm{xz,rms} \rangle$ close to the boundaries of the convective
layer, where the flow has to turn around, and in the stable layer filled with
internal gravity waves. The velocity profiles are nearly constant in the bulk of
the stable layer with $\langle \widetilde{v}_\mathrm{xz,rms} \rangle\,/\,\langle
\widetilde{v}_{y,\mathrm{rms}} \rangle\,{\approx}\,2$ and only a mild increase
in velocity towards lower densities. There is another drop in $\langle
\widetilde{v}_{y,\mathrm{rms}} \rangle$ at $y \gtrsim 2.9$ as gravity is turned
off (Eq.~\ref{eq:fg}), removing the waves' restoring force, and the upper wall
boundary condition forces $\langle \widetilde{v}_{y,\mathrm{rms}} \rangle$ to
vanish at $y = 3$. 

It is clear from Fig.~\ref{fig:velocity_profiles} that all \numcodes\ codes
produce similar time-averaged velocity profiles. However, we have to take the
stochastic character of turbulent convection into account to see whether the
remaining differences are significant. Instead of running ensembles of
simulations with randomised initial conditions, which would be rather expensive,
we obtain the $\pm 3 \sigma$ statistical-variation bands shown in
Fig.~\ref{fig:velocity_profiles} as follows. The central curve of each band
corresponds to the arithmetic average of all velocity profiles available on a
given computational grid. We also compute the standard deviations of each time
series in the averaging time window at each height $y$ and we average the
standard deviation profiles over the same set of runs to obtain one standard
deviation profile $\sigma_0(y)$. The profile $\sigma_0(y)$ is our best estimate
of the statistical variation to be expected in any of the runs, provided that
they are statistically similar, and it does not depend on any small systematic
differences between the velocity amplitudes predicted by different codes.
Obviously, the statistical variation associated with the time averages should
decrease as the length of the averaging interval is increased. We show above
that there is approximately one independent realisation of the convective flow
per turnover timescale $\tauconv$, so we estimate the statistical variation
associated with the time-averaged profiles to be $\sigma(y) =
\sigma_0(y)/\sqrt{N_\mathrm{conv}}$, where $N_\mathrm{conv}$ is the length of
the averaging time window in units of $\tauconv$.

However, we must keep in mind that $\sigma(y)$ is just an estimate, which
involves both statistical and systematic uncertainties. It depends on our
assumption that independent flow realisations are spaced by $\tauconv$ in time.
Figure~\ref{fig:autocorrelation} shows that the spacing could also be estimated
to be $0.5\tauconv$ or $2\tauconv$, depending on which set of runs we use and on
the very definition of `decorrelation'. Moreover, the autocorrelation function
$R(\Delta t)$ is based on the time evolution of the bulk convective
velocity. Different parts of the simulation domain may have different
characteristic timescales and our use of $\tauconv$ everywhere may bias the
estimate of $\sigma(y)$. We should thus use the statistical-variation bands as a
general guideline in quantifying differences between simulations but this simple
approach does not allow us to calculate the probability that a deviation of a
given magnitude is observed under the null hypothesis that the codes do not
differ.

Figure~\ref{fig:velocity_profiles} shows that the profiles of both velocity
components in the $128^3$ and $256^3$ runs fall within or close to the
respective estimated $\pm 3 \sigma$ statistical-variation bands. This means that
the small code-to-code differences are dominated by stochasticity. The bands of
both the $128^3$ and $256^3$ runs are slightly below the velocity curves of the
$512^3$ \ppmstar\ run in the convective layer, suggesting that velocity profiles
may not be fully converged on the $256^3$ grid. However, the time-averaging
window overlaps with the two above-mentioned episodes visible in
Fig.~\ref{fig:velocity_evolution}, during each of which the $512^3$ \ppmstar\
run reaches above-average bulk velocities for as much as ${\approx}\,2\tauconv$.
Given the uncertainty in determining the width of the statistical-variation
bands and the fact that we only have a single full-length $512^3$
run\footnote{The $512^3$ \prompi\ run is too short to be included in this
comparison, see Fig.~\ref{fig:velocity_evolution}.}, we do not find the tension
between the velocity profiles significant.

The arguments leading to our scaling the width of the statistical-variation
bands with $1/\sqrt{N_\mathrm{conv}}$ do not apply to the stable layer directly
because the response timescale of the stable layer is longer than $\tauconv$
(see above). Nevertheless, Fig.~\ref{fig:velocity_profiles} shows that
the statistical-variation bands describe the range of code-to-code variation
rather well. The velocity profiles of all $128^3$ and $256^3$ runs closely match
that of the $512^3$ \ppmstar\ run both in shape and amplitude, which further
supports our conclusion that the dominant wave patterns are essentially
converged already on the $128^3$ grid.

\begin{figure*}
\begin{minipage}{.5\textwidth}
  \centering
  \includegraphics[width=\textwidth]{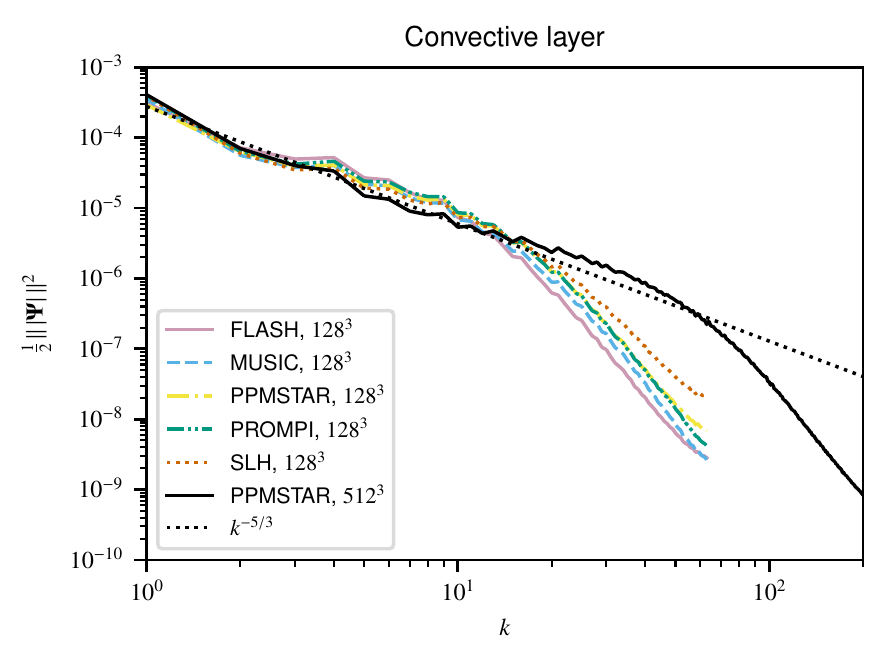}
\end{minipage}%
\begin{minipage}{.5\textwidth}
  \centering
  \includegraphics[width=\textwidth]{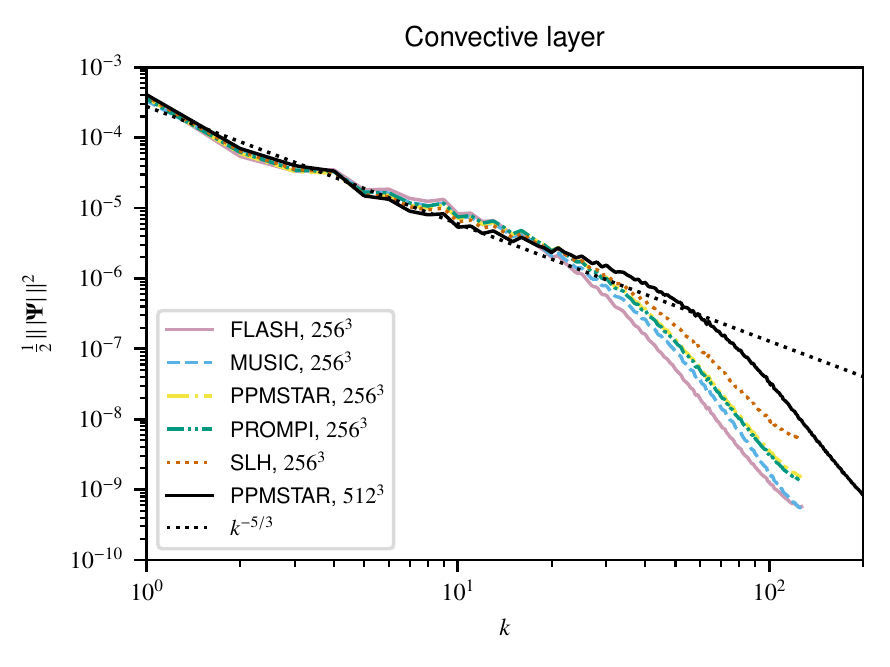}
\end{minipage}
\caption{Spectra of kinetic energy as functions of the wavenumber $k$ in a
   horizontal slice through the convective layer at $y = 1.7$. The spectra have
   been averaged over the whole analysis time interval ($500 \le t \le 2000$,
   approximately $19\,\tauconv$). The Kolmogorov scaling $k^{-5/3}$ is shown for
   comparison.}
\label{fig:spectra_cl}
\end{figure*}

The renderings of the vertical component of velocity $v_y$ in the horizontal
midplane of the convective layer presented in Fig.~\ref{fig:slices} show that
the convective flow is dominated by the largest possible scales with wavelengths
equal to the width of the computational box. This corresponds to convective
cells with an aspect ratio of about unity early in the simulations, which later
become slightly elongated in the vertical direction as the convective boundary
moves upwards. The cells are turbulent on smaller scales with large amounts of
small-scale vorticity $\omega$, also shown in Fig.~\ref{fig:slices} and in the
animations available on
Zenodo\footnote{\url{https://doi.org/10.5281/zenodo.5796842}}. We compute
spatial Fourier spectra of the velocity vector,
\begin{align}
   \bm{\Psi}_{jl} &= \frac{1}{N_x N_z} \sum_{m=0}^{N_x-1} \sum_{n=0}^{N_z-1}
   \bm{v}_{mn} \exp\left\{ -2 \pi i \left( \frac{m j}{N_x} + \frac{n l}{N_z}
   \right) \right\}, \\
   j &= 0, \dots, N_x - 1, \\
   l &= 0, \dots, N_z - 1,
   \label{eq:Fourier_transform}
\end{align}
where $N_x$ and $N_z$ are the total numbers of computational cells along the $x$
and $z$ axes, respectively. The velocity array $\bm{v}_{mn}$ corresponds to a
horizontal slice through the simulation box at $y = 1.7$, which is close to the
midplane of the convective layer at $t \approx 1250$. We compare the kinetic
energy per unit mass $\frac{1}{2} ||\, |\bm{\Psi}|\, ||^2$ binned over all
wavenumbers $k = \left( k_x^2 + k_z^2 \right)^{1/2}$, where
\begin{align}
   k_x &= 
      \begin{cases}
         j, & 0 \le j < \frac{N_x}{2} - 1,  \\
         -N_x + j, & \frac{N_x}{2} \le j < N_x, 
      \end{cases} \\
   k_z &= 
      \begin{cases}
         l, & 0 \le l < \frac{N_z}{2} - 1,  \\
         -N_z + l, & \frac{N_z}{2} \le l < N_z.
      \end{cases}
\end{align}
These expressions hold for even values of $N_x$ and $N_z$, which is the case for
all of our computational grids. Figure~\ref{fig:spectra_cl} shows the spectra
averaged in the whole time interval of analysis ($500 \le t \le 2000$). Although
the turbulence is anisotropic at small wavenumbers (large scales), it looks
close to being isotropic at the larger wavenumbers (smaller scales) that we see
in the renderings of vorticity in Fig.~\ref{fig:slices}. All of the codes
converge to the same kinetic energy spectrum upon grid refinement, which is
consistent with Kolmogorov's $-\frac{5}{3}$ law. Although we only use the $k_x$
and $k_z$ components of the wavenumber vector for simplicity, the kinetic-energy
spectrum of isotropic turbulence should have the same slope along all three axes
of the wavenumber space in the inertial range. If this was not the case there
would be more power along one or two of the axes on small scales in conflict
with the assumption of isotropy. 

The $512^3$ \ppmstar{} run illustrates that a rather fine grid is
needed to obtain a wide and well-converged inertial range. This is due to the
well-known bottleneck effect -- a power excess observed in numerical simulations
of turbulence between the inertial and dissipation ranges
\citep[e.g.][]{falkovich1994a, sytine2000a, dobler2003a}. All of our spectra
have similar shapes even in the dissipation range, although they diverge with
increasing $k$ and they reach a spread of as much as a factor of $10$ at the
Nyquist frequency. The spectrum in the dissipation range depends on the
behaviour of the numerical scheme close to the grid scale. However, this
dependence is largely irrelevant thanks to the fact that the dissipation rate
becomes independent of the magnitude of small-scale viscosity (be it of physical
or numerical origin) in turbulent flows as long as the viscosity is small enough.

\begin{figure*}
\begin{minipage}{.5\textwidth}
  \centering
  \includegraphics[width=\textwidth]{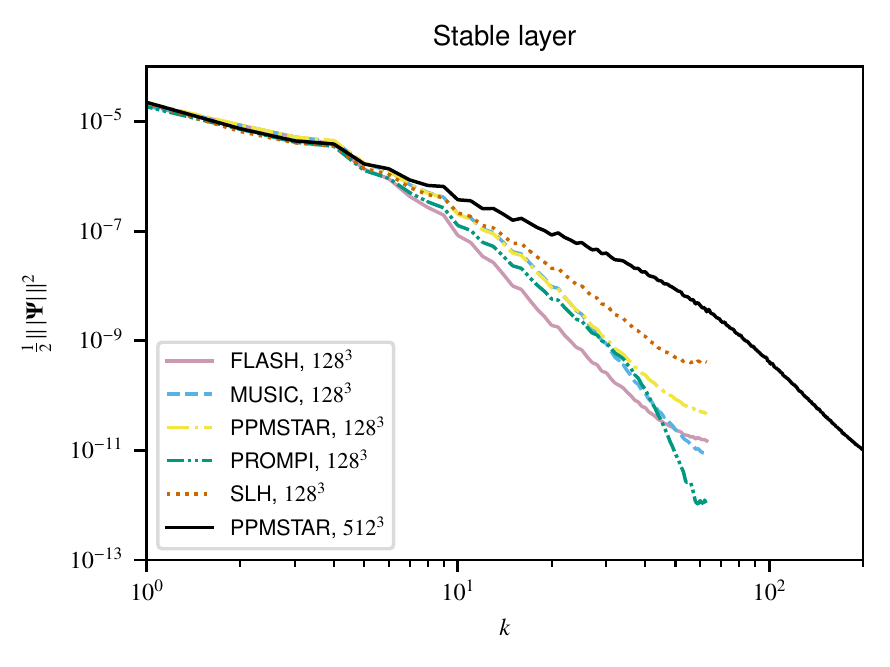}
\end{minipage}%
\begin{minipage}{.5\textwidth}
  \centering
  \includegraphics[width=\textwidth]{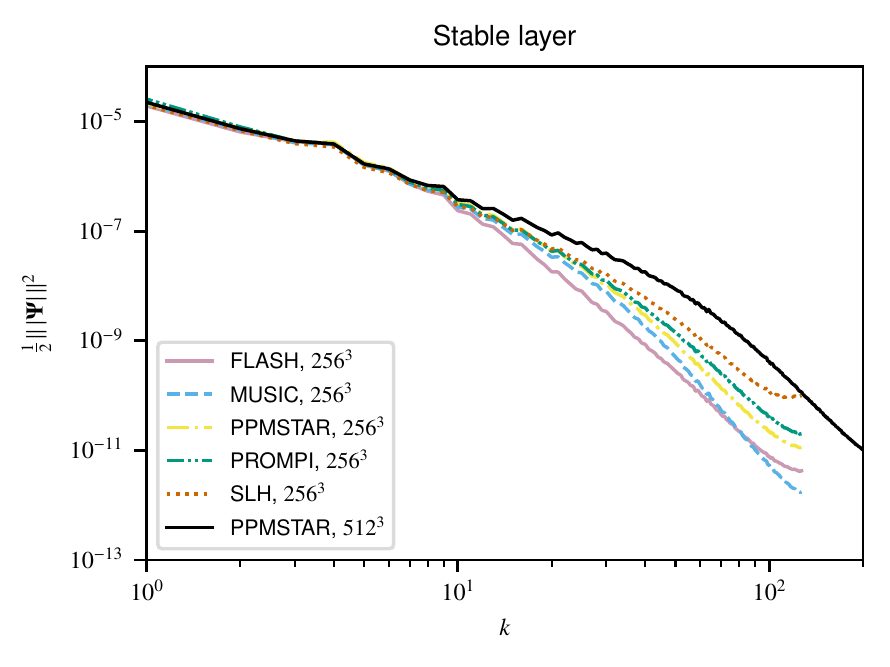}
\end{minipage}
\caption{Spectra of kinetic energy as functions of the wavenumber $k$ in a
horizontal slice through the stable layer at $y = 2.7$. The spectra have been
averaged over the whole analysis time interval ($500 \le t \le 2000$,
approximately $19\,\tauconv$).}
\label{fig:spectra_sl}
\end{figure*}

We use the same procedure to characterise the spatial spectra at $y = 2.7$ in
the stable layer, see Fig.~\ref{fig:spectra_sl}. These spectra are also
dominated by the largest scales, which can also be seen in the 2D renderings of
entropy\footnote{We refer to the pseudo-entropy $A = p/\rho^\gamma$ as
`entropy' for simplicity.} fluctuations in Fig.~\ref{fig:slices}. Although we
do not have any analytic prediction of the wave spectrum, all \numcodes{} codes
predict essentially the same spectrum at $k \lesssim 5$ on the $128^3$ grid and
at $k \lesssim 10$ on the $256^3$ grid, respectively. This corresponds to
horizontal wavelengths $\gtrsim 26$ computational cells in both cases. Although
this seems to be a large number, the actual challenge is to resolve the vertical
wavelength of the IGW, which is several times shorter as the waves are nearly
horizontal. The most extreme of these are revealed in the 2D renderings of
vorticity $\omega$ in Fig.~\ref{fig:slices}, which put more emphasis on shorter
vertical wavelengths as compared with the renderings of entropy fluctuations.

\subsection{Convective boundary and mass entrainment}
\label{sec:entrainment}

\begin{figure}
\includegraphics[width=\linewidth]{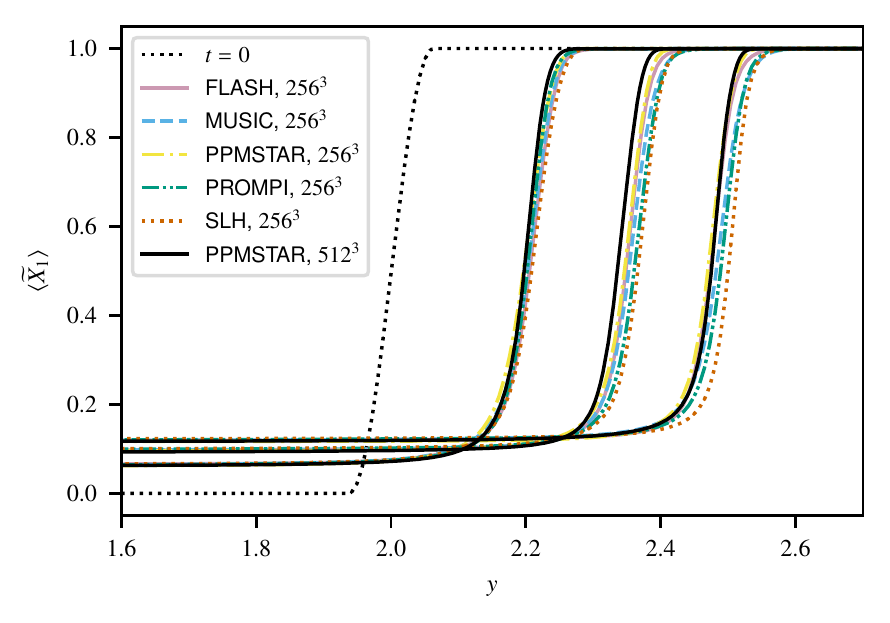}
\caption{Mass entrainment process visualised using vertical profiles of the
   mass fraction of the $\mu_1$ fluid. Only the uppermost parts of the
   convective layer and part of the stable layer are shown. The dotted line
   shows the initial condition and the remaining three sets of lines correspond
   to averages over time windows $\tauav \,=\, 2\,\tauconv \,=\, 160$ wide and
   centred on times of (from left to right) $580$, $1250$, and $1920$.}
\label{fig:X1_profiles_256}
\end{figure}

When upflows reach the upper boundary of the convective layer, they stir and
entrain some of the $\mu_1$ fluid in the transition zone and carry it down into
the convective layer as they turn around.\footnote{\citet{Woodward:15} provide a
detailed description of this process.} The distribution of this fluid's mass
fraction $X_1$, shown in Fig.~\ref{fig:slices}, provides a good visual
representation of the entrainment process. The process also involves the mixing
of entropy, which flattens the entropy gradient in the transition zone and makes
the convective layer grow. The gradual growth can be seen in the profiles of the
mass fraction $X_1$ shown in Fig.~\ref{fig:X1_profiles_256}.

\begin{figure*}
\begin{minipage}{.5\textwidth}
  \centering
  \includegraphics[width=\textwidth]{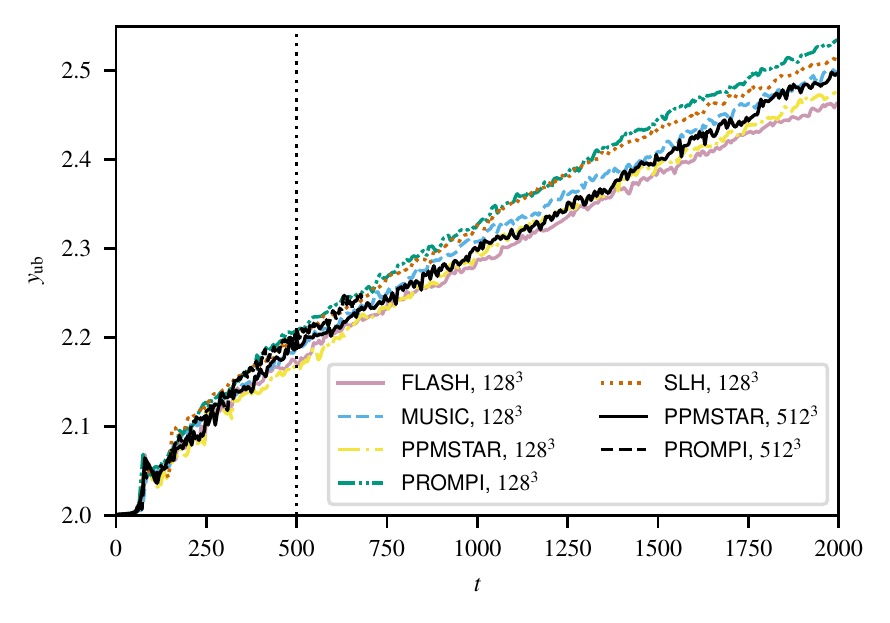}
\end{minipage}%
\begin{minipage}{.5\textwidth}
  \centering
  \includegraphics[width=\textwidth]{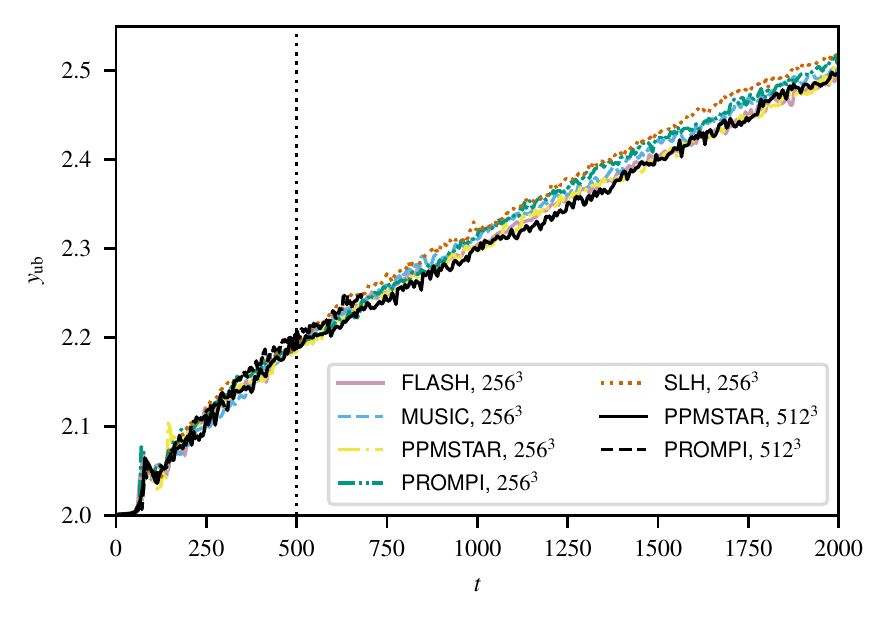}
\end{minipage}
\caption{Time evolution of the $y$ coordinate of the upper boundary of the
convection zone as determined from 1D averages. The dotted vertical line marks
the end of the initial transient excluded from the analysis.}
\label{fig:yub}
\end{figure*}

We define the vertical coordinate $y_\mathrm{ub}$ of the convective boundary to
be where $\partial_y \mw{X}_1$ reaches its global maximum. Discretisation noise
is reduced by fitting a parabola to the three data points closest to the
discrete estimate of the maximum's vertical coordinate. Figure~\ref{fig:yub}
shows that the $128^3$ runs diverge in $y_\mathrm{ub}(t)$ slightly with \flash{}
and \prompi{} predicting the slowest and fastest boundary motion, respectively.
However, the relative difference in the distance travelled by the end of the
simulations, $\Delta y_\mathrm{ub}(t_\mathrm{end}) =
y_\mathrm{ub}(t_\mathrm{end}) - y_\mathrm{ub}(0)$, is only ${\approx}14\%$
between these two extremes. The spread is reduced by another factor of
${\approx}\,3$ on the $256^3$ grid, on which all \numcodes{} codes agree on
$\Delta y_\mathrm{ub}(t_\mathrm{end})$ within $5\%$. Moreover, the
$y_\mathrm{ub}(t)$ curves derived from the $256^3$ runs closely track those from
the $512^3$ \prompi{} and \ppmstar{} runs.

\begin{figure*}
\begin{minipage}{.5\textwidth}
  \centering
  \includegraphics[width=\textwidth]{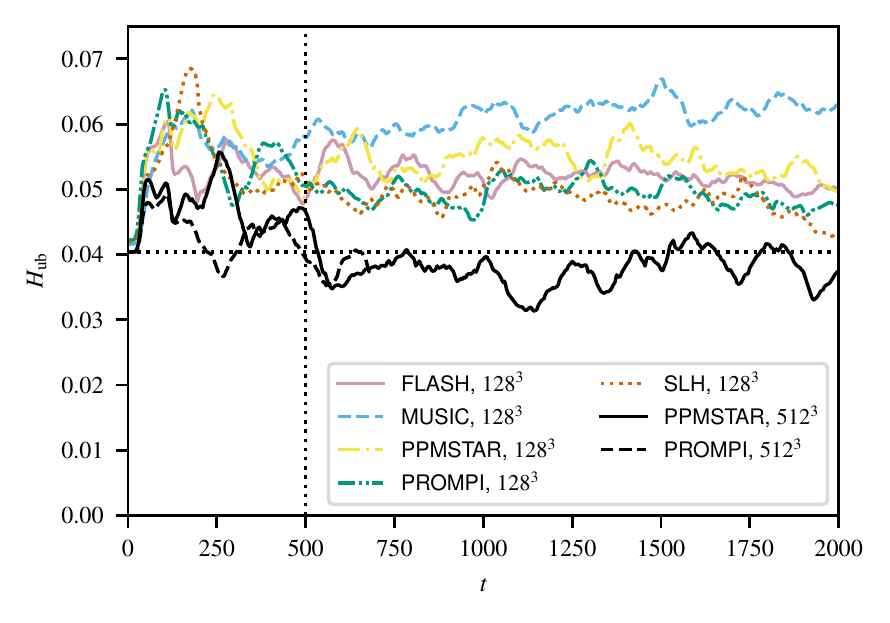}
\end{minipage}%
\begin{minipage}{.5\textwidth}
  \centering
  \includegraphics[width=\textwidth]{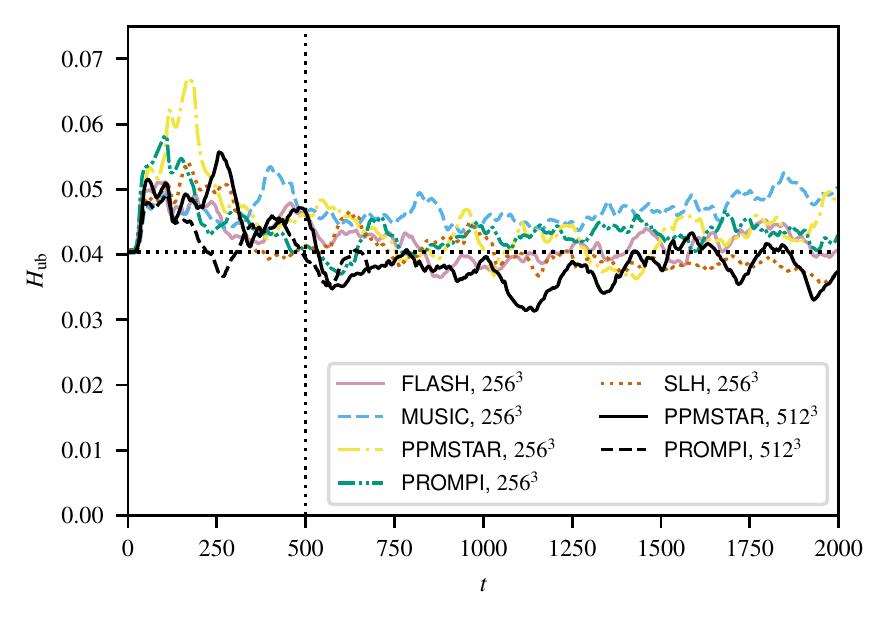}
\end{minipage}
\caption{Time evolution of the thickness $H_\mathrm{ub}$ (Eq.~\ref{eq:hub}) of
   the upper convective boundary. The time series have been smoothed using
   convolution with a top-hat kernel $\tauav \,=\, \tauconv \,=\, 80$ wide. The
   dotted vertical line marks the end of the initial transient excluded from the
   analysis. The dotted horizontal line shows $H_\mathrm{ub}(t\,{=}\,0) =
   0.0404$ for comparison.}
\label{fig:hub}
\end{figure*}

We characterise the thickness of the convective boundary using the scale height
\begin{equation}
   H_\mathrm{ub} = \left[ \mw{X}_1 \left( \frac{\partial
   \mw{X}_1}{\partial y} \right)^{-1} \right]_{y = y_\mathrm{ub}}.
   \label{eq:hub}
\end{equation}
Because this quantity is highly variable on short timescales, we smooth the
time series using a top-hat filter $\tauav\,{=}\,\tauconv\,{=}\,80$ wide and
show the result in Fig.~\ref{fig:hub}. In all \numcodes{} codes, the boundary is
up to $50\%$ thicker on the $128^3$ grid than on the $256^3$ grid. However,
differences between the $256^3$ and $512^3$ runs are substantially smaller,
suggesting that $H_\mathrm{ub}$ is close to being converged on the $256^3$ grid
at a value of ${\approx}\,0.04$. The converged value is close to the initial
value $H_\mathrm{ub}(t\,{=}\,0) =  0.0404$. This fact, however, is purely
coincidental. At $t\,{=}\,0$, $H_\mathrm{ub}$ characterises the steepness of the
1D transition zone as we define it. Once convection has started, $H_\mathrm{ub}$
is a product of spatial averaging along a 3D convective boundary that is as
sharp as the numerical scheme allows at some places but much wider at other
places, see Fig.~\ref{fig:slices}. The converged thickness
$H_\mathrm{ub}\,{\approx}\,0.04$ corresponds to $10$ computational cells on the
$512^3$ grid, although the PPB advection method implemented in \ppmstar{} can
preserve gradients spanning only about two computational cells
\citep{Woodward:15}. This suggests that the 3D deformation of the boundary
dominates the thickness of the spatially averaged boundary. However, it is much
more challenging for the codes to resolve the physical thickness of the boundary
on the $256^3$ and $128^3$ grids, on which $0.04$ units of length correspond to
only about $5$ and $2.5$ computational cells, respectively.

\begin{figure*}
\begin{minipage}{.5\textwidth}
  \centering
  \includegraphics[width=\textwidth]{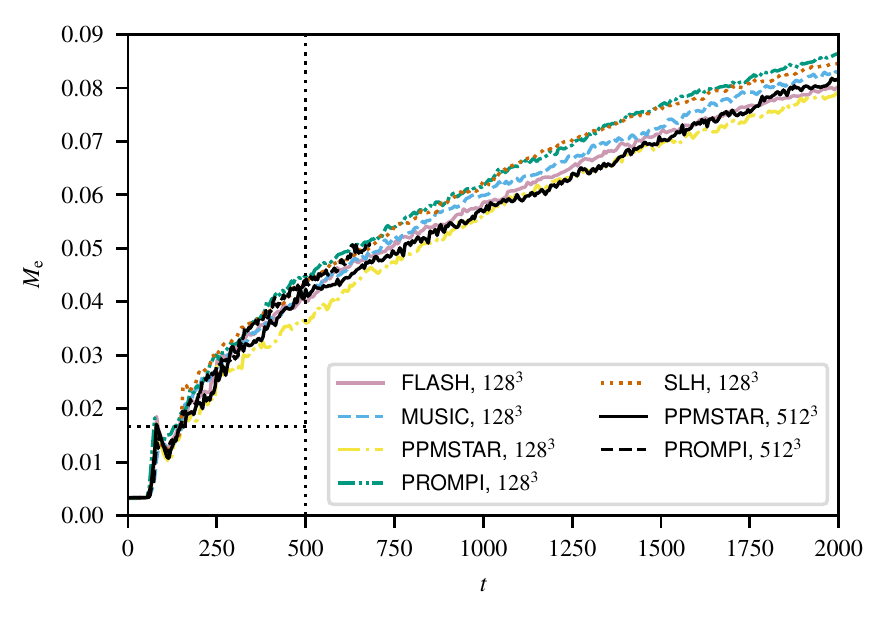}
\end{minipage}%
\begin{minipage}{.5\textwidth}
  \centering
  \includegraphics[width=\textwidth]{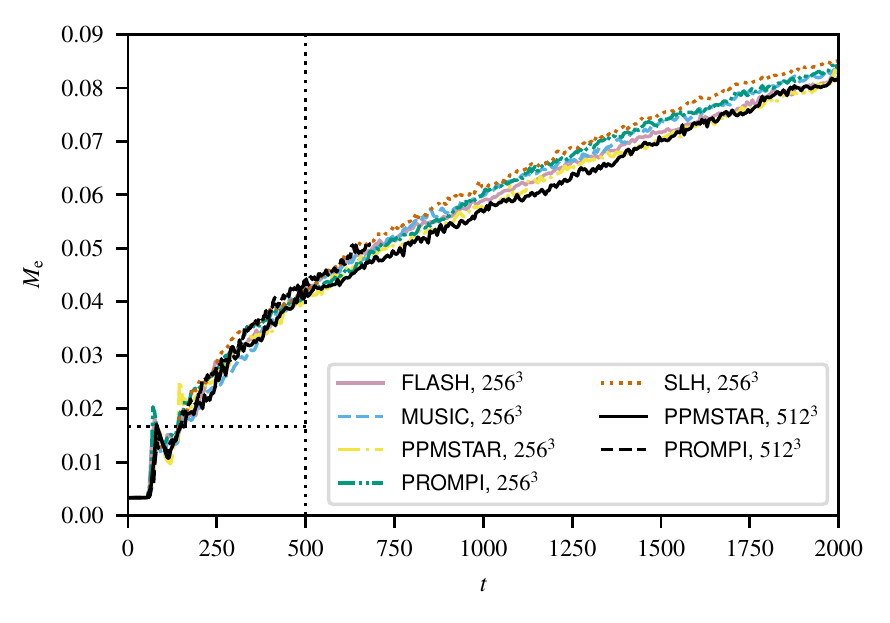}
\end{minipage}
\caption{Mass per unit horizontal area of the $\mu_1$ fluid entrained into the
   convective layer as a function of time. The dotted vertical line marks the
   end of the initial transient excluded from the analysis. The dotted
   horizontal line shows the amount of the $\mu_1$ fluid contained in the
initial transition zone between the convective and stable layer.}
\label{fig:me}
\end{figure*}

The total amount of the $\mu_1$ fluid entrained by time $t$ is
\begin{equation}
   M_\mathrm{e}(t) = \int_{1}^{y_\mathrm{ub}(t)} \mw{X}_1(y,t)\,
   \vw{\rho}(y,t)\, \mathrm{d}y
\end{equation}
per unit of horizontal surface area and is shown in Fig.~\ref{fig:me}. This
entrainment metric differs slightly from $y_\mathrm{ub}(t)$ because of the
density stratification. The agreement between the codes' predictions is slightly
better as compared with the $y_\mathrm{ub}(t)$ metric and the $128^3$ and
$256^3$ simulations agree within $9\%$ and $4\%$, respectively, on the total
amount of the $\mu_1$ fluid entrained by the end of the simulations. All of the
initial transition layer is entrained early on during the initial transient as
indicated in Fig.~\ref{fig:me}.

\begin{figure*}
\begin{minipage}{.5\textwidth}
  \centering
  \includegraphics[width=\textwidth]{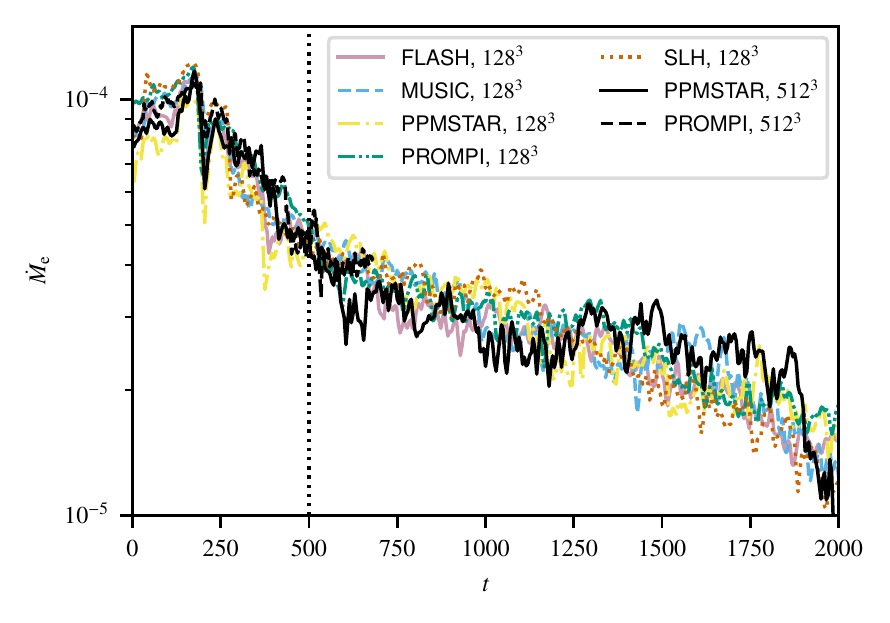}
\end{minipage}%
\begin{minipage}{.5\textwidth}
  \centering
  \includegraphics[width=\textwidth]{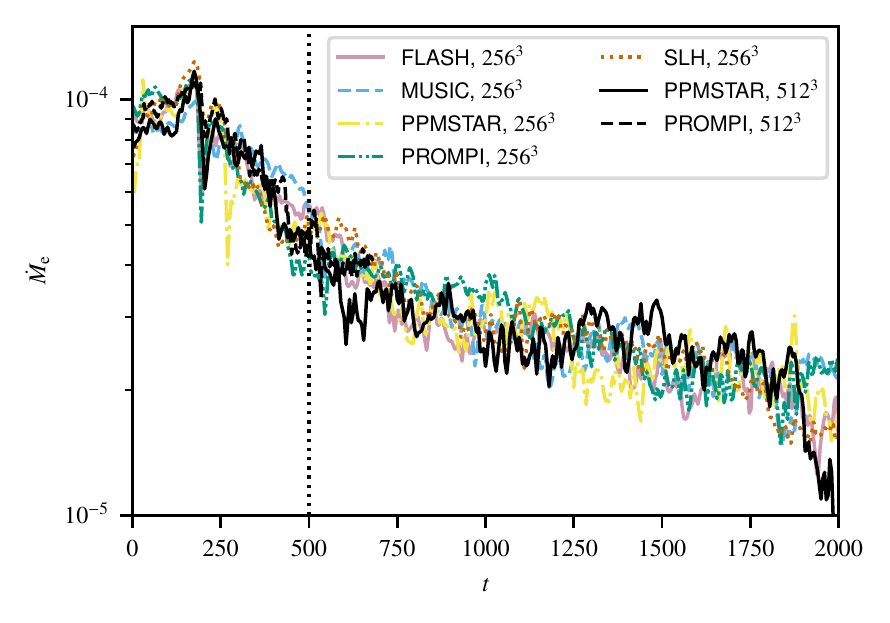}
\end{minipage}
\caption{Mass entrainment rates corresponding to the curves shown in
Fig.~\ref{fig:me}. The time series have been smoothed using convolution with a
top-hat kernel $\tauav = 3\,\tauconv = 240$ wide. The dotted vertical line marks
the end of the initial transient excluded from the analysis.}
\label{fig:dmedt}
\end{figure*}

It is also clear from Fig.~\ref{fig:me} that the mass entrainment rate
$\dot{M}_\mathrm{e}(t)$ is relatively high early on and slowly decreases
throughout the simulation time. We compute $\dot{M}_\mathrm{e}(t)$ using
second-order central differences. The differencing greatly amplifies noise,
which we suppress using convolution with a centred top-hat kernel
$\tauav\,{=}\,3\tauconv\,{=}\,240$ wide. The mass entrainment rates, shown in
Fig.~\ref{fig:dmedt}, agree between all of the codes within their statistical
fluctuations already on the $128^3$ grid and they remain unchanged as the grid
is refined up to $512^3$. However, $\dot{M}_\mathrm{e}(t)$ varies randomly on
timescales as long as several convective turnover timescales, which likely
contributes to the small spread in $M_\mathrm{e}(t)$ in Fig.~\ref{fig:me}. The
entrainment rate decreases from ${\approx}\, 1.0 \times 10^{-4}$ at the
beginning of the initial transient to ${\approx}\, 4.5 \times 10^{-5}$ at $t =
t_0$ and to ${\approx}\, 2.0 \times 10^{-5}$ at $t = t_\mathrm{end}$. Part of
the decrease may be attributed to the density at the convective boundary, which
decreases from $0.28$ at $t = 0$ to $0.15$ at $t = t_\mathrm{end}$.

One may find it surprising that the mass entrainment rates obtained on the
coarse $128^3$ grid agree so well with one another. Perhaps even more
surprisingly, they also agree with the rates obtained in the $512^3$ \prompi\
and \ppmstar\ runs. Convective mass entrainment is usually thought of as a
complex process sensitive to small-scale flows and instabilities in the boundary
layer. However, \citet{spruit2015a} argues that the mass entrainment rate is set
by the amount of energy available to overcome the buoyancy of the fluid being
entrained; this was also observed in a laboratory study by \citet{linden1975a}.
Applying this constraint in models of stars with helium-burning convective cores
improves their agreement with both asteroseismology and population studies of
globular clusters \citep{constantino2017}. The entrainment rate is then
proportional to the luminosity $L$ driving the convective flow as confirmed by
the 3D simulations of \citet{jones2017a} and \citet{andrassy2020a} of oxygen
burning in a setup similar to our present test setup. More evidence comes from
calibrations of the entrainment law $\dot{M}_\mathrm{e} \propto v_\mathrm{rms}\,
\mathrm{Ri_{B}}^{-n}$, where $v_\mathrm{rms}$ is the rms velocity of convection
and $\mathrm{Ri_{B}}$ is the bulk Richardson number proportional to
$v_\mathrm{rms}^{-2}$ for a given convective boundary \citep[for a complete
definition, see][]{MeakinArnett2007}. Assuming that $v_\mathrm{rms}
\,{\propto}\, L^{1/3}$, we have $\dot{M}_\mathrm{e} \,{\propto}\, L^{(1 +
2n)/3}$ and $n \,{=}\, 1$ corresponds to $\dot{M}_\mathrm{e} \,{\propto}\, L$.
Values of $n$ measured in different numerical simulations range from $0.74 \pm
0.04$ \citep{cristini2019a} and $0.74 \pm 0.01$ \citep{horst2021a} through $1.05
\pm 0.21$ \citep{MeakinArnett2007} to $1.32 \pm 0.79$ \citep{higl21a}. If
buoyancy is the dominant factor one can expect to obtain a good estimate of the
entrainment rate as soon as the largest downflows are reasonably resolved.

Finally, the very fact that we keep increasing the mean entropy of the
convective layer by heating it at the bottom contributes to mass entrainment.
This process, previously mentioned by \citet{MeakinArnett2007},
\citet{andrassy2020a}, and \citet{horst2021a}, can be understood as follows. The
convective layer is well mixed and its entropy is essentially constant in space.
On the other hand, entropy increases with height in the stable layer
(Fig.~\ref{fig:stratification}). If the mean entropy of the convective layer
increases it becomes higher than that of a thin layer at the very bottom of the
stable layer. This thin layer is thus negatively buoyant and it must sink and
mix with the convective layer. In our simulations, we compute the rate of change
of the mean entropy in the lower $\nicefrac{2}{3}$ of the convective layer to
eliminate any influence of the convective boundary. The value of $\mathrm{d} A /
\mathrm{d}t$ reaches $4.7 \times 10^{-5}$ by $t \approx 150$ in our $256^3$ runs
and it remains statistically constant until the end of the simulations. The
entrainment rate due to the process described above is
$\dot{M}_\mathrm{e,heating} = (\mathrm{d} A / \mathrm{d}t) \,/\, (\mathrm{d} A /
\mathrm{d}m_1)$, where $m_1$ is the cumulative mass of the $\mu_1$ fluid
integrated from the bottom of the simulation box upwards. We measure the entropy
gradient $\mathrm{d} A / \mathrm{d}m_1$ in the initial stratification. Its value
is $1.6$ at the top of the initial transition layer between the two fluids and
$3.8$ where $m_1 = M_\mathrm{e}(t = t_\mathrm{end}) \doteq 0.083$ (see
Fig.~\ref{fig:me}). This implies that $\dot{M}_\mathrm{e,heating} = 2.9 \times
10^{-5}$ early on during the initial transient, which is $\approx 30\%$ of the
entrainment rate $\dot{M}_\mathrm{e}$ measured. This contribution increases to
as much as $\approx 60\%$ by the end of the simulations. 

\subsection{Fluctuations and transport of energy}
\label{sec:fluctuations_fluxes}

\begin{figure*}
\begin{minipage}{.5\textwidth}
  \centering
  \includegraphics[width=\textwidth]{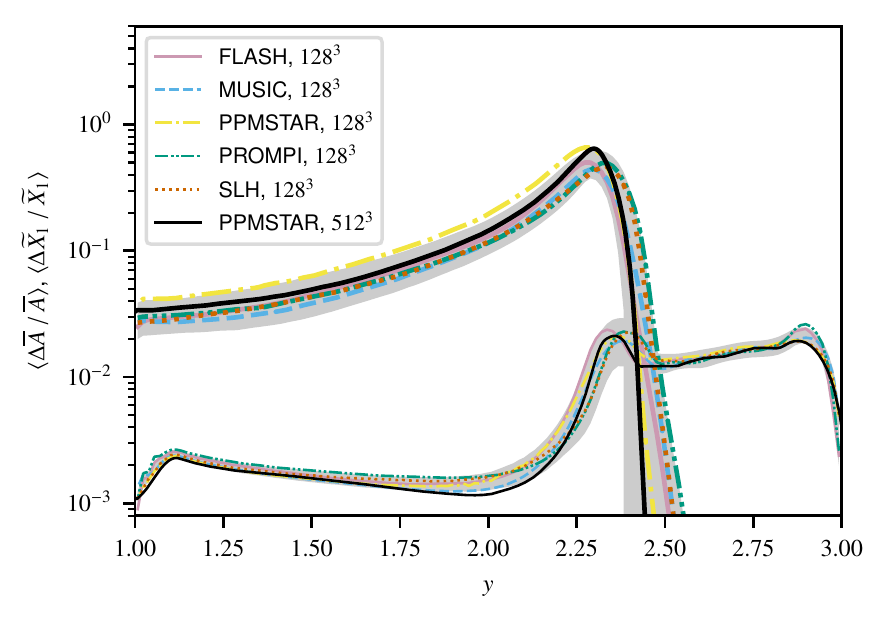}
\end{minipage}%
\begin{minipage}{.5\textwidth}
  \centering
  \includegraphics[width=\textwidth]{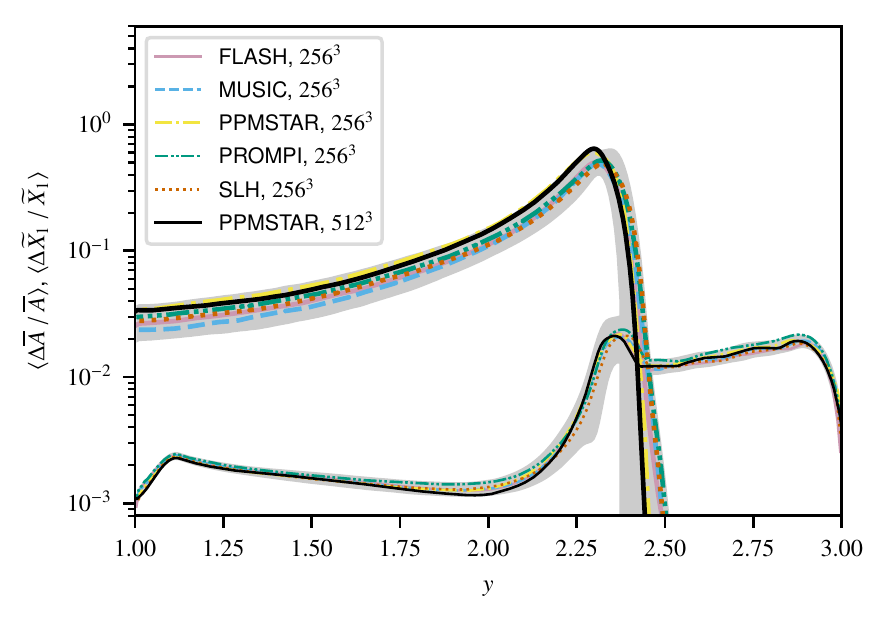}
\end{minipage}
\caption{Relative fluctuations in entropy $\vw{A}$ (thin lines) and mass
   fraction $\mw{X}_1$ of the $\mu_1$ fluid (thick lines) averaged in a time
   interval \mbox{$\tauav = 6\,\tauconv = 480$} wide and centred on time $1250$.
   The grey bands give an estimate of $\pm 3 \sigma$ statistical variation in
   the averages due to stochasticity, see Sect.~\ref{sec:velocity_field} for
   details.}
\label{fig:fluctuations}
\end{figure*}

The rates of convective transport of energy and chemical composition scale with
the flow speed and with the magnitude of fluctuations in the flow field. We have
already shown that the flow speed is code-independent, see
Sect.~\ref{sec:velocity_field}. Figure~\ref{fig:fluctuations} shows that the
same holds for the time-averaged relative fluctuations in entropy $\langle
\Delta \vw{A}/\vw{A} \rangle$ and in the mass fraction of the $\mu_1$ fluid
$\langle \Delta \vw{X}_1/\vw{X}_1 \rangle$. The figure includes statistical-variation
bands computed as described in Sect.~\ref{sec:velocity_field}. The profiles of
the fluctuations as produced by all of the \numcodes\ codes fall within or close
to the corresponding statistical-variation band. The profiles of $\langle \Delta
\vw{A}/\vw{A} \rangle$ agree better on the $256^3$ grid in the bulk of the
convection zone, although the differences are small already on the $128^3$ grid.
The slight code-to-code differences in the time-averaged position of the
convective boundary (where $\langle \Delta \vw{X}_1/\vw{X}_1 \rangle$ drops)
decrease as the mass entrainment rate converges upon grid refinement, see
Sect.~\ref{sec:entrainment}.

We define the flux of enthalpy to be the quantity \begin{equation} \mathcal{F}_H
= \vw{H v_y} - \vw{H}\, \mw{v}_y, \label{eq:fh} \end{equation} where $H$ is
enthalpy per unit volume and $v_y$ is the vertical component of velocity. The
second term in Eq.~\ref{eq:fh} is included to remove the flux contribution of
the thermal expansion and compression of the horizontally averaged
stratification. We average the enthalpy flux over all of the simulation time
after the initial transient ($500 \le t \le 2000$), ignoring the upward
propagation of the convective boundary and focusing on the bulk of the
convective layer instead. The profiles of $\mathcal{F}_H(y)$ shown in
Fig.~\ref{fig:fh_profiles} agree well within the statistical-variation bands
computed as described in Sect.~\ref{sec:velocity_field}, although our method
seems to overestimate the statistical variation in this quantity as it is
significantly larger than the code-to-code differences.

\begin{figure}
\includegraphics[width=\linewidth]{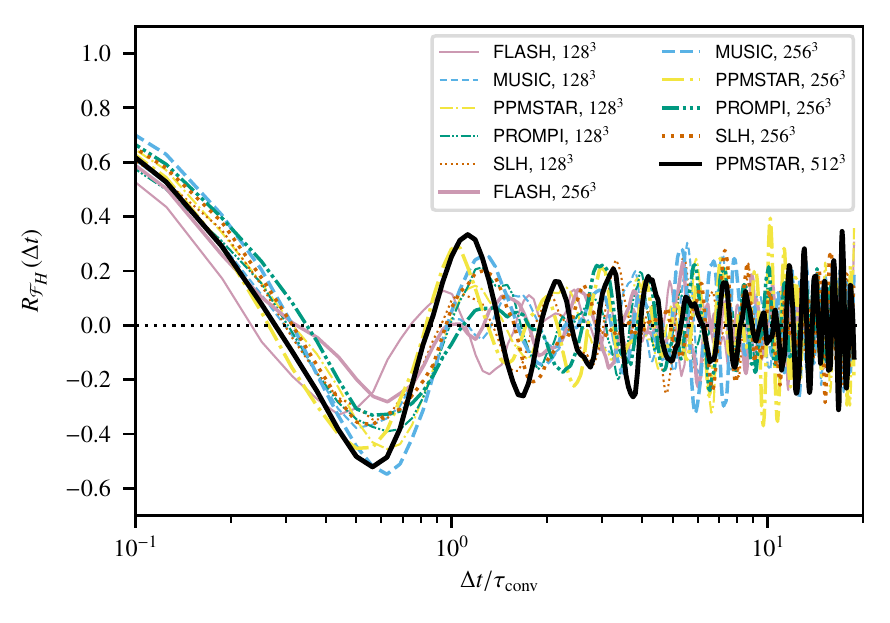}
\caption{Autocorrelation function $R_{\mathcal{F}_H}(\Delta t)$ of
$\mathcal{F}_H(y \,{=}\, 1.5)$ as a function of the time shift $\Delta t$ in
units of the convective turnover timescale $\tauconv$.}
\label{fig:autocorrelation_FH}
\end{figure}

\begin{figure*}
\begin{minipage}{.5\textwidth}
  \centering
  \includegraphics[width=\textwidth]{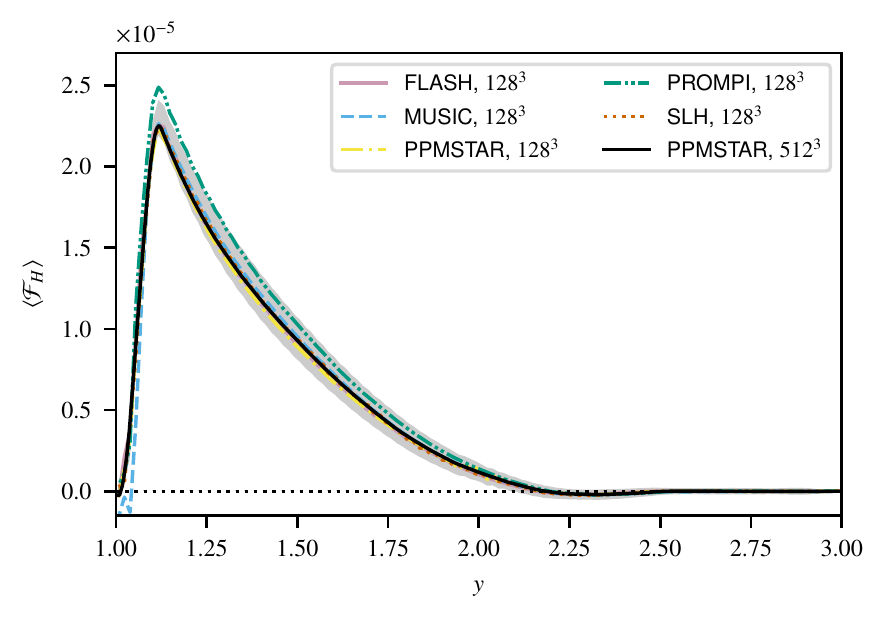}
\end{minipage}%
\begin{minipage}{.5\textwidth}
  \centering
  \includegraphics[width=\textwidth]{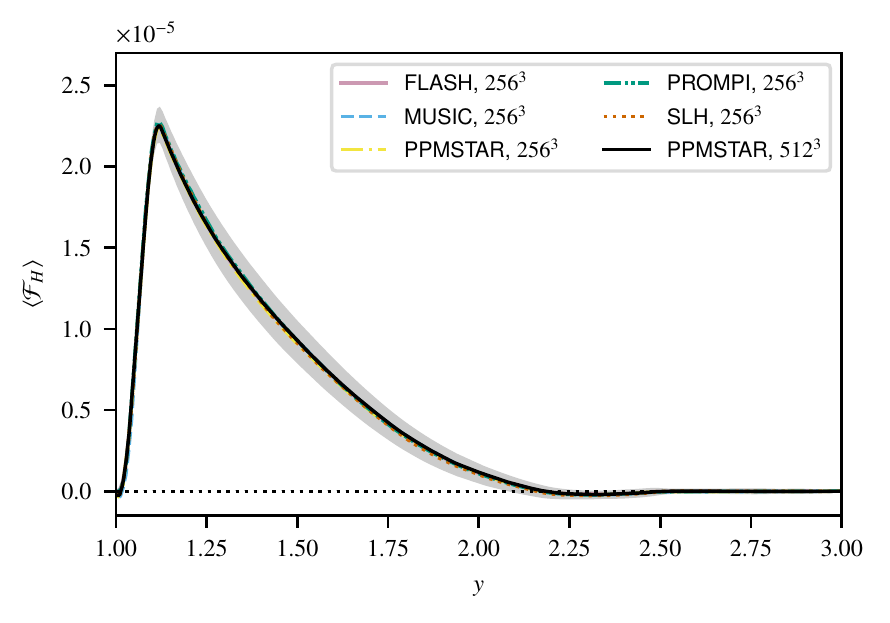}
\end{minipage}
\caption{Vertical profiles of the enthalpy flux averaged over the whole analysis
time interval ($500 \le t \le 2000$, approximately $19\,\tauconv$). The grey
bands give an estimate of $\pm 3 \sigma$ statistical variation in the averages due to
stochasticity, see Sect.~\ref{sec:velocity_field} for details.}
\label{fig:fh_profiles}
\end{figure*}

To explain this, we compute the autocorrelation function
$R_{\mathcal{F}_H}(\Delta t)$ of $\mathcal{F}_H(y \,{=}\, 1.5)$ using the method
described in Sect.~\ref{sec:velocity_field}. The shape of
$R_{\mathcal{F}_H}(\Delta t)$ turns out to be rather different from the
autocorrelation function $R(\Delta t)$ of the bulk convective velocity; compare
Fig.~\ref{fig:autocorrelation_FH} with Fig.~\ref{fig:autocorrelation}. First,
the autocorrelation drops to zero on a timescale as short as ${\approx}\,
0.3\tauconv$. This suggests that the width of the statistical variation bands
should be reduced by the factor $0.3^{-\frac{1}{2}} = 1.8$, see
Sect.~\ref{sec:velocity_field}. Moreover, all of our runs show some
anticorrelation for $0.3\tauconv \lesssim \Delta t \lesssim \tauconv$ before the
values of $R_{\mathcal{F}_H}(\Delta t)$ start to oscillate around zero for
longer time shifts. The anticorrelation suppresses fluctuations in long-term
time averages. From the physical point of view, the anticorrelation may reflect
the fact that changes to the heat flux divergence result in local heating or
cooling of the nearly isentropic stratification and the convective instability
provides strong negative feedback, quickly driving the flux profile back to its
quasi-stationary shape. In light of this, the ${\approx}10\%$ overestimation of
$\mathcal{F}_H(y)$ in the lower convection zone in the $128^3$ \prompi\ run may
be statistically significant. However, the $256^3$ \prompi\ run agrees well with
other codes run on the same grid as well as with the $512^3$ \ppmstar\ run. The
small differences between the codes as well as the possible feedback mechanism
suggest that $\mathcal{F}_H$ is a poor indicator of code or simulation quality.

\begin{figure*}
\begin{minipage}{.5\textwidth}
  \centering
  \includegraphics[width=\textwidth]{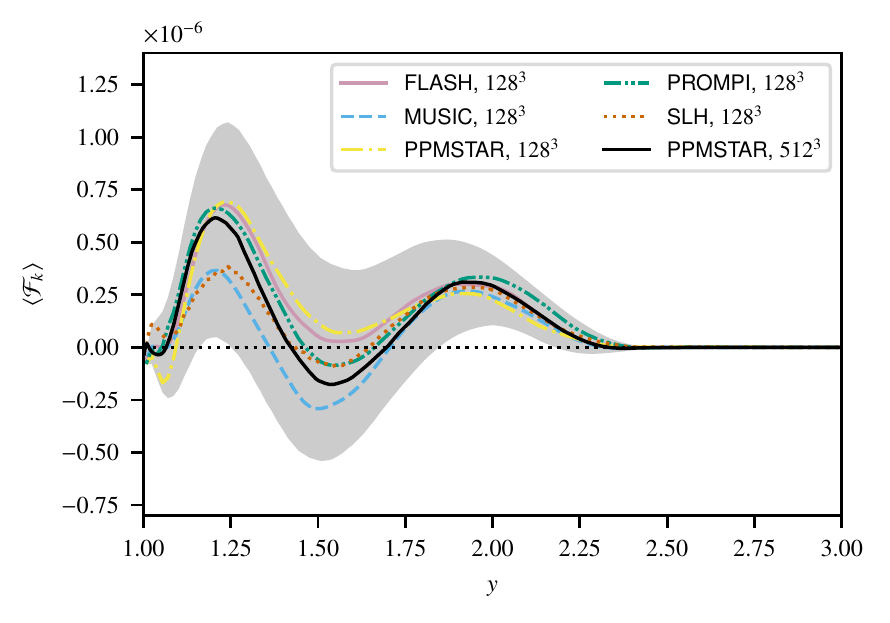}
\end{minipage}%
\begin{minipage}{.5\textwidth}
  \centering
  \includegraphics[width=\textwidth]{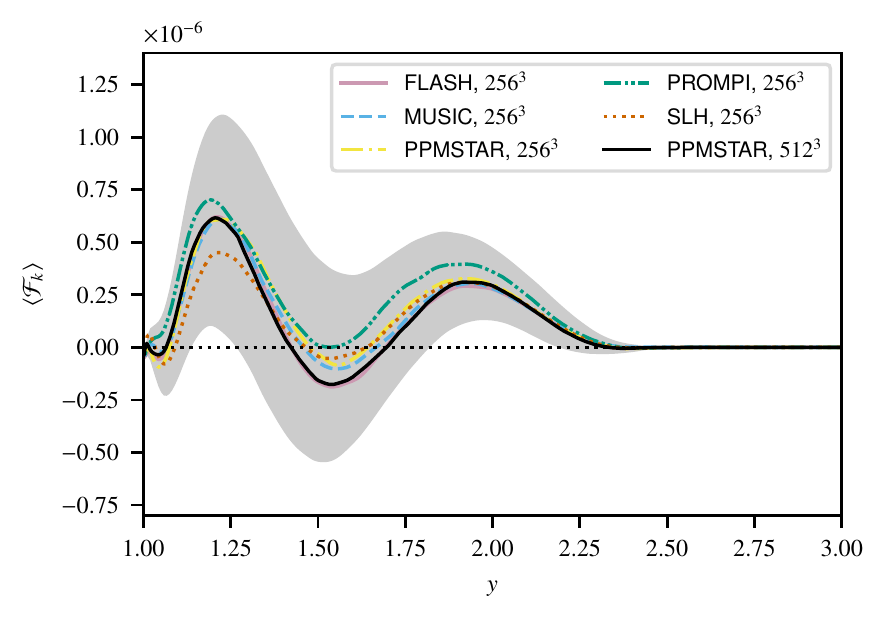}
\end{minipage}
\caption{Vertical profiles of the kinetic-energy flux averaged over the whole
   analysis time interval ($500 \le t \le 2000$, approximately $19\,\tauconv$).
   The grey bands give an estimate of $\pm 3 \sigma$ statistical variation in
   the averages due to stochasticity, see Sect.~\ref{sec:velocity_field} for
   details.}
\label{fig:fk_profiles}
\end{figure*}

The flux of kinetic energy is defined in a way analogous to Eq.~\ref{eq:fh},
\begin{equation}
   \mathcal{F}_\mathrm{k} = \frac{1}{2}\left( \vw{\rho\, |\bm{v}|^2\, v_y}
   - \vw{\rho\, |\bm{v}|^2}\, \mw{v}_y \right).
   \label{eq:fk}
\end{equation}
Figure~\ref{fig:fk_profiles} shows that the amplitude of
$\mathcal{F}_\mathrm{k}$ is as much as ${\approx}\,30$ times smaller than that of
$\mathcal{F}_H$ in our simulations. Stochasticity introduces large statistical
variation into the time-averaged profiles of this small quantity. Nevertheless,
all of the $128^3$ and $256^3$ runs agree with each other as well as with the
$512^3$ \ppmstar\ run on the profile of $\mathcal{F}_\mathrm{k}$ within the
statistical-variation bands included in the figure.

\begin{figure*}
\begin{minipage}{.5\textwidth}
  \centering
  \includegraphics[width=\textwidth]{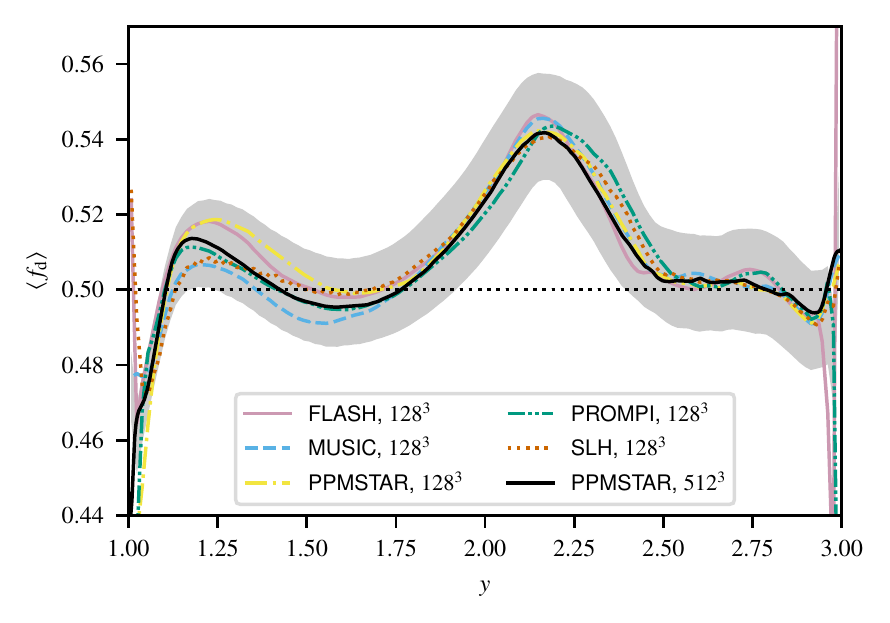}
\end{minipage}%
\begin{minipage}{.5\textwidth}
  \centering
  \includegraphics[width=\textwidth]{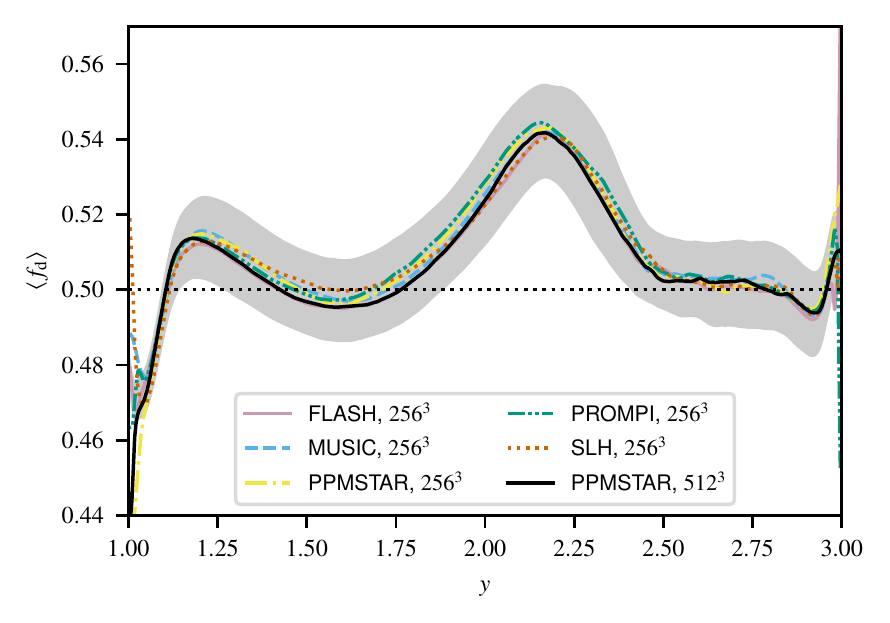}
\end{minipage}
\caption{Vertical profiles of the downflow filling factor averaged over the
   whole analysis time interval ($500 \le t \le 2000$, approximately
   $19\,\tauconv$). The grey bands give an estimate of $\pm 3 \sigma$
   statistical variation in the averages due to stochasticity, see
Sect.~\ref{sec:velocity_field} for details.}
\label{fig:ffd_profiles}
\end{figure*}

Why is the magnitude of $\mathcal{F}_\mathrm{k}$ so much smaller than that of
$\mathcal{F}_H$? Using a simplified, two-stream model of convection, we show in
Appendix~\ref{sec:fk_two_stream} that the magnitude of $\mathcal{F}_\mathrm{k}$
depends on the degree of asymmetry between upflows and downflows. We
characterise the asymmetry using a downflow filling factor $f_\mathrm{d}$
defined to be the relative horizontal area covered by flows with vertical
velocity $v_y - \mw{v}_y < 0$. In the two-stream model, $\mathcal{F}_\mathrm{k}$
vanishes for $f_\mathrm{d} = 0.5$, that is when there is perfect up-down
symmetry. Figure~\ref{fig:ffd_profiles} shows that $f_\mathrm{d}$ indeed is
close to $0.5$ in our simulations. In contrast to $\mathcal{F}_\mathrm{k}$,
$\mathcal{F}_H$ remains a substantial fraction of the total flux even if there
is perfect up-down symmetry because convection must transport heat across the
convective layer. This explains why the amplitude of $\mathcal{F}_\mathrm{k}$ is
much smaller than that of $\mathcal{F}_H$ in our simulations. The two-stream
model also predicts that $\mathcal{F}_\mathrm{k}$ is negative (i.e. directed
downwards) for $f_\mathrm{d} < 0.5$ and positive for $f_\mathrm{d} > 0.5$. We
observe the same trend in our simulations: compare Figs.~\ref{fig:fk_profiles}
and \ref{fig:ffd_profiles} and see also Fig.~\ref{fig:fk_profiles_two_stream}.
The role of $\mathcal{F}_\mathrm{k}$ is much more important in strongly
stratified, convective stellar envelopes, in which downflows tend to be
substantially narrower than upflows and $\mathcal{F}_\mathrm{k}$ is large and
negative (i.e. directed downwards). \citet{VialletMeakin2013} show this
difference directly in their Fig.~13, in which they compare their 3D
hydrodynamic models of the convective envelope of a red giant and of an
oxygen-burning shell in a massive star. The latter model is rather similar to
our test setup and, not surprisingly, they also find that the amplitude of
$\mathcal{F}_\mathrm{k}$ is ${\approx}\,30$ times smaller than that of
$\mathcal{F}_H$ in that model (compare their Figs.~13 and 15).

\section{Summary and conclusions}
\label{sec:summary}

In this work, we have defined a test problem involving adiabatic turbulent
convection and mass entrainment from a stably stratified layer. The problem is
directly relevant to current 3D simulation efforts to model convection in
stellar interiors. The problem's physics and geometry are simplified to make it
accessible to a wide range of 3D hydrodynamic codes but to retain the crucial
processes we are investigating. Because turbulent flows are involved, this test
problem is fundamentally different from standard test problems. There is no
single solution and the simulations must be compared in a statistical way,
taking care not to overinterpret statistical fluctuations as real differences
between codes or simulations. We compare simulations computed using the codes
\flash, \music, \ppmstar, \prompi, and \slh, which we run for $25$ convective
turnover timescales to gather as much statistics as possible.

All of the simulations, computed on Cartesian grids ranging from $128^3$ through
$256^3$ to $512^3$ cells, show a turbulent convective layer dominated by
large-scale flows with an rms Mach number of ${\approx}\,0.04$. Upflows turning
around at the upper convective boundary entrain mass from the bottom of the
overlying stable layer and pull it into the convective layer. They also generate
internal gravity waves, which are present throughout the stable layer. The bulk
rms velocities as well as time- and space-averaged velocity profiles in both the
convective and stable layer are statistically the same in all \numcodes\ codes
even on the coarse $128^3$ grid. They also agree with the profiles obtained on
finer grids, although the $512^3$ \ppmstar\ run deviates slightly because of two
long episodes of increased velocity, likely of statistical origin. All of the
codes converge to the same kinetic energy spectrum upon grid refinement, which
is consistent with Kolmogorov's $-\frac{5}{3}$ scaling. The \numcodes\ codes
slightly differ in their dissipation rates close to the grid scale, which is
reflected in slight differences in the shape of the kinetic energy spectrum in
the dissipation range. The codes also give essentially the same spectrum in the
stable layer for horizontal wavelengths $\gtrsim 26$ computational cells.

The convective boundary is somewhat under-resolved on the $128^3$ grid, being as
much as ${\approx}\,50\%$ thicker than its converged thickness value as
determined by the $256^3$ and $512^3$ runs. Nevertheless, the simulations agree
on the total amount of mass entrained by the simulations' end within $9\%$
already on the $128^3$ grid. This improves to $4\%$ on the $256^3$ grid and the
remaining spread seems to be dominated by stochasticity. The rapid convergence
of the mass entrainment rate upon grid refinement is compatible with the idea of
\citet{spruit2015a} that the rate is set by the global energetics of the whole
process as opposed to details of the small-scale instabilities occurring in the
boundary layer. We also show that approximately $30$ to $60$ per cent of the
entrainment rate can be attributed to a process in which mass entrainment is
caused by the constant heating of the convective layer and the subsequent
increase in its mean entropy.

Finally, all of the codes statistically agree on the time-averaged profiles of
fluctuations in composition and entropy and on the profiles of the enthalpy- and
kinetic-energy fluxes. This agreement is somewhat better on the $256^3$ grid as
compared to the $128^3$ one. We give likely reasons for the rapid convergence of
the time-averaged enthalpy flux, which makes it a rather poor indicator of code
or simulation quality. The flux of kinetic energy is ${\approx}\,30$ times
smaller than the flux of enthalpy in our test problem. We explain this disparity
using a toy model showing that the flux of kinetic energy is very small indeed
when the horizontal areas covered by upflows and downflows are close to being
the same as observed in our simulations.

All in all, we find excellent code-to-code agreement on a rather complex
turbulent-flow problem. The numerical schemes implemented in these codes differ
in many aspects such as numerical flux functions, reconstruction and
time-stepping methods, or the use (or not) of dimensional splitting. It would
certainly be interesting to compare our results with those obtained using finite
difference or spectral methods, which are not included in the present work. To
facilitate future studies of this kind, we formulated the test problem in
Cartesian geometry and such that the Mach number of the flows (${\approx}\,0.04$
rms) is easy to reach for a wide range of codes. We provide our data as well as
all of the setup files and analysis scripts needed if anyone should be
interested in performing and analysing simulations of the kind presented here in
the future, see Appendix~\ref{sec:supplement}. A useful, if quite expensive,
future extension of this study could involve a series of simulations to see how
weakly one can drive the convection and still maintain reasonable code-to-code
agreement on affordable grids. With a lower driving luminosity, the mass
entrainment rate would also decrease and allow us to gather statistics of the
waves in the stable layer on long timescales, which would provide well-resolved
temporal wave spectra and test the accuracy of the codes' asteroseismic
predictions.

\begin{acknowledgements}
PVFE was supported by the U.S. Department of Energy through the Los Alamos
National Laboratory (LANL). LANL is operated by Triad National Security, LLC,
for the National Nuclear Security Administration of the U.S. Department of
Energy (Contract No. 89233218CNA000001). This work has been assigned a document
release number LA-UR-21-25840. RA, JH, LH, GL, and FKR acknowledge support by
the Klaus Tschira Foundation. The work of FKR is supported by the German
Research Foundation (DFG) through the grant RO~3676/3-1. This work is funded by
the Deutsche Forschungsgemeinschaft (DFG, German Research Foundation) under
Germany's Excellence Strategy EXC 2181/1 - 390900948 (the Heidelberg STRUCTURES
Excellence Cluster). The authors gratefully acknowledge the Gauss Centre for
Supercomputing e.V. (www.gauss-centre.eu) for funding this project by providing
computing time through the John von Neumann Institute for Computing (NIC) on the
GCS Supercomputer JUWELS \citep{JUWELS} at Jülich Supercomputing Centre (JSC).
FH acknowledges funding through an NSERC Discovery Grant. This work has
benefitted from and motivated in part by research performed as part of the JINA
Center for the Evolution of the Elements (NSF Grant No. PHY-1430152). PRW
acknowledges support from NSF grants 1814181, 2032010, and PHY-1430152, as well
as computing support through NSF's Frontera computing system at TACC. The
software used in this work was in part developed by the DOE NNSA-ASC OASCR
\flash\ Center at the University of Chicago. This work is partly supported by
the ERC grant No. 787361- COBOM and the consolidated STFC grant ST/R000395/1.
The authors would like to acknowledge the use of the University of Exeter
High-Performance Computing (HPC) facility ISCA. DiRAC Data Intensive service at
Leicester, operated by the University of Leicester IT Services, which forms part
of the STFC DiRAC HPC Facility. The equipment was funded by BEIS capital funding
via STFC capital grants ST/K000373/1 and ST/R002363/1 and STFC DiRAC Operations
grant ST/R001014/1. RH acknowledges support from the World Premier International
Research Centre Initiative (WPI Initiative), MEXT, Japan and the IReNA AccelNet
Network of Networks, supported by the National Science Foundation under Grant
No. OISE-1927130. This article is based upon work from the ChETEC COST Action
(CA16117), supported by COST (European Cooperation in Science and Technology).
This project has received funding from the European Union’s Horizon 2020
research and innovation programme under grant agreement No 101008324
(ChETEC-INFRA). This work used the DiRAC@Durham facility managed by the
Institute for Computational Cosmology on behalf of the STFC DiRAC HPC Facility
(www.dirac.ac.uk). The equipment was funded by BEIS capital funding via STFC
capital grants ST/P002293/1 and ST/R002371/1, Durham University, and STFC
operations grant ST/R000832/1. This work also used the DiRAC Data Centric system
at Durham University, operated by the Institute for Computational Cosmology on
behalf of the STFC DiRAC HPC Facility. This equipment was funded by BIS National
E Infrastructure capital grant ST/ K00042X/1, STFC capital grants ST/H008519/1
and ST/ K00087X/1, STFC DiRAC Operations grant ST/K003267/1, and Durham
University. DiRAC is part of the National E Infrastructure. SWC acknowledges
federal funding from the Australian Research Council through a Future Fellowship
(FT160100046) and Discovery Project (DP190102431). This work was supported by
computational resources provided by the Australian Government through NCI via
the National Computational Merit Allocation Scheme (project ew6), and resources
provided by the Pawsey Supercomputing Centre which is funded by Australian
Government and the Government of Western Australia. We thank the anonymous
referee for constructive comments that improved this paper.
\end{acknowledgements}

\bibliographystyle{aa}
\bibliography{ccpaper}

\begin{appendix}

\section{Supplementary materials}
\label{sec:supplement}

To make this study easy to reproduce or extend, we provide our data products as
well as setup and analysis scripts in electronic form. The 1D and 2D data,
animations and spectra are available on
Zenodo\footnote{\url{https://doi.org/10.5281/zenodo.5796842}}. A set of Jupyter
notebooks and Python scripts that can set up the test problem and reproduce our
analysis as well as all plots shown in this work can be found on the
\mbox{CoCoPy} GitHub
repository\footnote{\url{https://github.com/robert-andrassy/CoCoPy}}, v1.1.0.
The analysis makes use of the PyPPM
toolkit\footnote{\url{https://github.com/PPMstar/PyPPM}}, v2.0.1. To make the
analysis even more approachable, we set up the CoCo Jupyter
Hub\footnote{\url{https://www.ppmstar.org/coco}}, which contains all of the data
and scripts needed. The hub also contains the much more voluminous original
$128^3$ and $256^3$ data sets. We will consider adding the $512^3$ runs upon
request.

The hydrodynamic codes write 3D output into binary data files, the structure of
which is code-dependent. We have created a Python interface for each of the
codes. The interface makes sure that the 3D data are read and assigned to a set
of arrays, which are then further processed in a code-independent way. The
processing is performed in the system of units defined in Table~\ref{tab:units}
independently of what units the hydrodynamic codes use. As detailed below, we
produce text files containing 1D horizontal averages and kinetic energy spectra
as well as binary files containing 2D slices through the simulation box. This
intermediate step speeds up data visualisation and facilitates data sharing. 

The 1D horizontal averages are written into text-based \texttt{.rprof}
files\footnote{The extension \texttt{.rprof} is of historical origin: such files
have mostly been used to store radial profiles of full-sphere \ppmstar\
simulations of stellar convection.} with one file per each output interval. The
files contain a two-line header followed by a set of data tables containing the
quantities summarised in Table~\ref{tab:rprof_vars} such that the files are both
easily human- and machine-readable.

The kinetic-energy spectra are computed in the $y \,{=}\, 1.7$ and $y \,{=}\,
2.7$ planes using Eq.~\ref{eq:Fourier_transform}. They are written into
text-based \texttt{.spec} files with one file per each output interval. The
files contain a one-line header followed by two data tables, the first for the
$y \,{=}\, 1.7$ plane and the other for the $y \,{=}\, 2.7$ plane. The tables
list the value of $\frac{1}{2} ||\, |\bm{\Psi}|\, ||^2$ as a function of the
wavenumber $k$ ranging from $k \,{=}\, 0$ to the Nyquist wavenumber.

At each output interval, we also produce a set of \texttt{NumPy}'s binary
\texttt{.npy} files, each containing one variable in a 2D slice through the
computation box. The slicing planes are $x \,{=}\, 0$, $y \,{=}\, 1.7$, $y
\,{=}\, 2.7$, and $z \,{=}\, 0$ and the variables are the pseudo-entropy $A
\,{=}\, p / \rho^\gamma$, density $\rho$, mass fraction $X_1$ of the $\mu_1$
fluid, the Cartesian components of velocity $v_x$, $v_y$, and $v_z$, magnitude
of vorticity $|\bm{\nabla} \times \bm{v}|$, and velocity divergence $\bm{\nabla}
\cdot \bm{v}$.

\begin{table*}[h]
\caption{Definitions of the horizontally averaged quantities available in the
\texttt{.rprof} files.}
\label{tab:rprof_vars}
\centering
\begin{tabular}{lp{40mm}p{42.5mm}l}
   \toprule
   Variable & Meaning & Definition & Statistics \\
   Name     &         &            &            \\
   \midrule
   \code{Y} & vertical coordinate & $y$ & \ldots \\
   \code{RHO} & density & $\vw{\rho}$ & full \\
   \code{P} & pressure & $\vw{p}$ & full \\
   \code{TEMP} & temperature & $\vw{T}$ & full \\
   \code{A} & pseudo-entropy & $\vw{A};\ A = p / \rho^\gamma$ & full \\
   \code{X1} & mass fraction of the lighter fluid & $\mw{X_1}$ & full \\
   \code{V} & magnitude of velocity & $\mw{|\bm{v}|}$ & full \\
   \code{VX} & $x$-component of velocity & $\mw{v}_x$ & full \\
   \code{VY} & $y$-component of velocity & $\mw{v}_y$ & full \\
   \code{VZ} & $z$-component of velocity & $\mw{v}_z$ & full \\
   \code{\textbar VY\textbar} & magnitude of $v_y$ & $\mw{|v_y|}$ & average \\
   \code{VXZ} & magnitude of horizontal velocity & $\mw{v}_{xz};\
   v_{xz} = \sqrt{v_x^2 + v_z^2}$ & average \\
   \code{VORT} & magnitude of vorticity & $\mw{\omega};\ \omega = |\bm{\nabla}
   \times \bm{v}|$ & average \\
   \code{FK} & flux of kinetic energy & $\mathcal{F}_\mathrm{k} = \frac{1}{2}\,
   (\vw{\rho |\bm{v}|^2 v_y} - \vw{\rho |\bm{v}|^2}\, \mw{v}_y)$ & \ldots \\
   \code{FH} & flux of enthalpy & $\mathcal{F}_H = \vw{H v_y} - \vw{H} \mw{v_y}$
   & \ldots \\
   \code{FFD} & downflow filling factor & fraction of the total horizontal area
   where $v_y - \mw{v}_y < 0$ & \ldots \\
   \bottomrule
\end{tabular}
\vspace{0.5em}
\tablefoot{In the last column, `average' means that only the average is
   computed and `full' indicates that also the minimum, maximum, and standard
   deviation are computed. The names of the last three quantities then start
   with \texttt{MIN\_}, \texttt{MAX\_}, and \texttt{STDEV\_}, respectively.}
\end{table*}

\section{Flux of kinetic energy in a two-stream model of convection}
\label{sec:fk_two_stream}

In this section, we derive a simple relation between the asymmetry between
upflows and downflows, quantified by the downflow filling factor, and the flux
of kinetic energy in a 1D, two-stream model of convection. Although purely
kinematic in nature, the model shows how contributions from the upflows cancel
those from the downflows, making the flux of kinetic energy vanish when the flow
is close to perfect up-down symmetry.

\begin{figure}
\includegraphics[width=\linewidth]{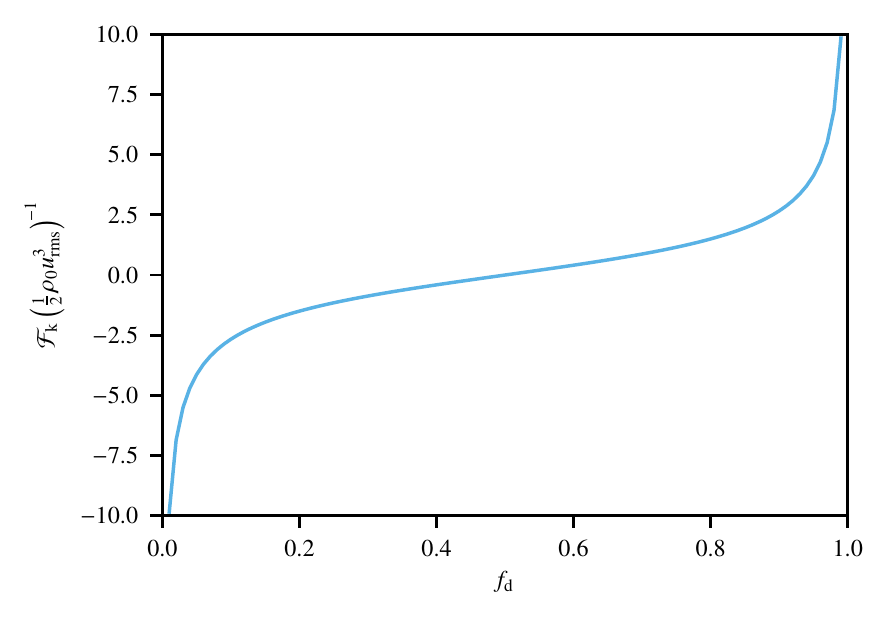}
\caption{Geometric factor $\mathcal{F}_\mathrm{k} \left( \frac{1}{2} \rho_0
u_\mathrm{rms}^3 \right)^{-1}$ in Eq.~\ref{eq:fk_two_stream} as a function of
the downflow filling factor $f_\mathrm{d}$.}
\label{fig:fk_two_stream}
\end{figure}

We consider a 1D model of convection that consists of an upflow and a downflow
such that the net mass flux through the whole horizontal plane vanishes. Both
streams are assumed to have the same density $\rho_0$ for
simplicity.\footnote{We can neglect density fluctuations because (1) we do not
model the dynamics of convection here and (2) the fluctuations are small.} Let
\mbox{$0 < f_\mathrm{d} < 1$} be the geometrical filling factor (relative
surface area) of the downflow. The net mass flux is then
\begin{equation}
\rho_0 (1 - f_\mathrm{d}) u_\mathrm{u} + \rho_0 f_\mathrm{d} u_\mathrm{d} = 0,
\end{equation}
where $u_\mathrm{u} > 0$ and $u_\mathrm{d} < 0$ are the velocities of the upflow
and downflow, respectively. The ratio of the velocity components is
\begin{equation}
\frac{u_d}{u_u} = -\frac{1 - f_\mathrm{d}}{f_\mathrm{d}}
\label{eq:velocity_ratio}
\end{equation}
and the rms velocity is
\begin{equation}
u_\mathrm{rms} = \sqrt{(1 - f_\mathrm{d}) u_\mathrm{u}^2 + f_\mathrm{d}
u_\mathrm{d}^2}.
\end{equation}
Using Eq.~\ref{eq:velocity_ratio}, we can express the individual velocities of
the two streams in terms of $u_\mathrm{rms}$:
\begin{align}
   u_\mathrm{u} &= \sqrt{\frac{f_\mathrm{d}}{1 - f_\mathrm{d}}}
   u_\mathrm{rms},\\
   u_\mathrm{d} &= -\sqrt{\frac{1 - f_\mathrm{d}}{f_\mathrm{d}}} u_\mathrm{rms}.
\end{align}
The net flux of kinetic energy is
\begin{equation}
\mathcal{F}_\mathrm{k} = \frac{1}{2} \rho_0 \left[ (1 - f_\mathrm{d})
u_\mathrm{u}^3 + f_\mathrm{d} u_\mathrm{d}^3 \right],
\label{eq:fk_two_stream0}
\end{equation}
which can be expressed as a function of $f_\mathrm{d}$ and $u_\mathrm{rms}$:
\begin{equation}
\mathcal{F}_\mathrm{k} = \frac{1}{2} \rho_0 \,u_\mathrm{rms}^3\, f_\mathrm{d}
\left(\frac{1 - f_\mathrm{d}}{f_\mathrm{d}}\right)^\frac{3}{2} \left[
\left(\frac{f_\mathrm{d}}{1 - f_\mathrm{d}}\right)^2 - 1 \right].
\label{eq:fk_two_stream}
\end{equation}
Figure~\ref{fig:fk_two_stream} shows the geometric factor
$\mathcal{F}_\mathrm{k} \left( \frac{1}{2} \rho_0 u_\mathrm{rms}^3
\right)^{-1}$. Because $\rho_0 > 0$ and $u_\mathrm{rms} > 0$, the flux is
negative (i.e. directed downwards) for $f_\mathrm{d} < 0.5$, positive for
$f_\mathrm{d} > 0.5$, and it vanishes when $f_\mathrm{d} = 0.5$, that is when
there is perfect up-down symmetry and the upflow and downflow contributions
cancel each other in Eq.~\ref{eq:fk_two_stream0}.

\begin{figure}
\includegraphics[width=\linewidth]{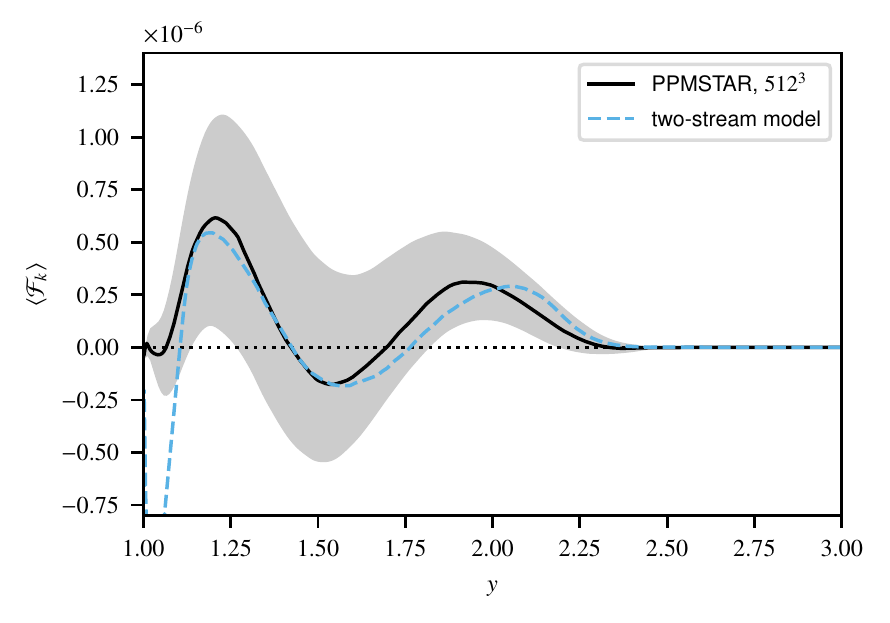}
\caption{Profile of the kinetic-energy flux $\mathcal{F}_\mathrm{k}$ given by
   the two-stream model as compared with the actual profile of
   $\mathcal{F}_\mathrm{k}$ in the $512^3$ \ppmstar\ run. Inputs for the
   two-stream model are taken from the same \ppmstar\ run. The time-averaging
interval and statistical-variation bands are the same as those used in
Fig.~\ref{fig:fk_profiles}.}
\label{fig:fk_profiles_two_stream}
\end{figure}

We test the two-stream model by taking its input quantities from the $512^3$
\ppmstar\ run and comparing the result with the actual profile of
$\mathcal{F}_\mathrm{k}$ in the same run. In particular, we set $\rho_0 =
\langle \vw{\rho} \rangle$, $f_\mathrm{d} = \langle f_\mathrm{d} \rangle$ and
$u_\mathrm{rms}^3 = \langle \mw{v}_\mathrm{rms} \rangle^2 \, \langle
\mw{v}_{y,\mathrm{rms}} \rangle$ in Eq.~\ref{eq:fk_two_stream}. The last
expression is motivated by the fact that we are computing the flux of kinetic
energy along the $y$ axis. We use the whole analysis time interval ($500 \le t
\le 2000$, approximately $19\,\tauconv$) for the time averaging.
Figure~\ref{fig:fk_profiles_two_stream} shows that the two-stream model closely
reproduces the actual profile of $\mathcal{F}_\mathrm{k}$ both in magnitude and
sign over most of the convective layer.

\end{appendix}
\end{document}